\documentclass[aoas]{imsart}
\usepackage{color}
\usepackage{epsfig}
\usepackage{amsfonts}
\usepackage{amsmath}
\usepackage{amssymb, amsbsy}
\usepackage{multicol}
\usepackage{rotating}
\usepackage{multirow}
\usepackage{chngpage}
\RequirePackage{natbib}
\usepackage{url}
\startlocaldefs
\DeclareMathOperator{\var}{var}
\DeclareMathOperator{\se}{se}
\DeclareMathOperator{\dopearl}{do}
\endlocaldefs
\defcitealias{ehret2011}{Ehret et al., 2011}
\defcitealias{dastani2012}{Dastani et al., 2012}
\defcitealias{johnson2013vig}{Johnson, 2013}
\begin{document}
\begin{frontmatter}
\title{Integrating summarized data from multiple genetic variants in Mendelian randomization: bias and coverage properties of inverse-variance weighted methods}
\runtitle{Summarized data in Mendelian randomization: bias and coverage}

\author{\fnms{Stephen} \snm{Burgess} \corref{} \ead[label=e1]{sb452@medschl.cam.ac.uk}}
\address{Strangeways Research Laboratory \\ 2 Worts Causeway \\ Cambridge, CB1 8RN \\ UK \\ \printead{e1}}
\affiliation{Department of Public Health and Primary Care, University of Cambridge}

\author{\fnms{Jack} \snm{Bowden} \ead[label=e2]{jack.bowden@bristol.ac.uk}}
\address{Oakfield House \\ Oakfield Grove \\ Bristol, BS8 2BN \\ UK \\ \printead{e2}}
\affiliation{MRC Integrative Epidemiology Unit, University of Bristol}

\runauthor{Stephen Burgess and Jack Bowden}

\begin{abstract}
Mendelian randomization is the use of genetic variants as instrumental variables to assess whether a risk factor is a cause of a disease outcome. Increasingly, Mendelian randomization investigations are conducted on the basis of summarized data, rather than individual-level data. These summarized data comprise the coefficients and standard errors from univariate regression models of the risk factor on each genetic variant, and of the outcome on each genetic variant. A causal estimate can be derived from these associations for each individual genetic variant, and a combined estimate can be obtained by inverse-variance weighted meta-analysis of these causal estimates. Various proposals have been made for how to calculate this inverse-variance weighted estimate. In this paper, we show that the inverse-variance weighted method as originally proposed (equivalent to a two-stage least squares or allele score analysis using individual-level data) can lead to over-rejection of the null, particularly when there is heterogeneity between the causal estimates from different genetic variants. Random-effects models should be routinely employed to allow for this possible heterogeneity. Additionally, over-rejection of the null is observed when associations with the risk factor and the outcome are obtained in overlapping participants. The use of weights including second-order terms from the delta method is recommended in this case.
\end{abstract}

\begin{keyword}
\kwd{Mendelian randomization}
\kwd{instrumental variables}
\kwd{summarized data}
\kwd{weak instruments}
\kwd{causal inference}
\end{keyword}

\end{frontmatter}

\section{Introduction}
Mendelian randomization is the use of genetic variants as instrumental variables to investigate the causal effect of a modifiable risk factor on an outcome using observational data \citep{burgess2015book}. Mendelian randomization analyses are increasingly performed using summarized data, rather than individual-level data \citep{burgess2014twosample}. There are various methods for combining the estimates from multiple genetic variants into a single causal estimate \citep{burgess2013genepi}. In particular, an inverse-variance weighted method has been proposed \citep{johnson2013vig} that is equivalent (for a particular choice of weights) to the standard two-stage least squares method usually employed with individual-level data \citep{burgess2014pleioajeappendix}. However, different authors have used different formulae for estimating the variances of the estimates that are used as weights \citep{dastani2012, shen2015}. Additionally, some authors have used fixed-effect meta-analysis for the combination of estimates from different genetic variants \citep{nelson2015}, whereas other authors have used random-effects meta-analysis \citep{ahmad2015}.

In this paper, we compare the bias and coverage properties of estimates from the inverse-variance weighted method for different choices of weights, and using fixed-effect, additive random-effects, and multiplicative random-effects models for combining the estimates. In Section~\ref{sec:meth}, we introduce the inverse-variance weighted method, and demonstrate its equivalence to both a two-stage least squares analysis and to a weighted linear regression of the association estimates. We also present the different versions of the method that are investigated further in this paper. In Section~\ref{sec:motivate}, we provide an example analysis that was the motivation for this work. In this example, subtly different choices in the analysis method result in estimates that differ considerably and lead to substantively different conclusions. In Section~\ref{sec:simulate}, we perform a simulation study to compare the bias and coverage properties of the different versions of the method. Finally, in Section~\ref{sec:discuss}, we discuss the findings of this paper and their relevance to applied practice.

\section{Methods}
\label{sec:meth}
We provide a brief introduction to Mendelian randomization -- the use of genetic variants as instrumental variables; further introductory references to the subject area are available \citep{daveysmith2003, lawlor2007, schatzkin2009}. The objective of Mendelian randomization is to judge whether intervention on a modifiable risk factor would affect a disease outcome. This is achieved by testing whether genetic variants that satisfy the assumptions of an instrumental variable for the risk factor are associated with the outcome. An instrumental variable is a variable that is associated with the risk factor, but not associated with confounders of the risk factor--outcome association, nor is there any causal pathway from the instrumental variable to the outcome except for that via the risk factor (see \cite{greenland2000, martens2006} for further information on instrumental variables). This means that the genetic variant is an unconfounded proxy for variation in the risk factor, and therefore can be treated as similar to treatment assignment in a randomized trial, where the treatment is to change the level of the risk factor \citep{nitsch2006}. Similarly to an intention-to-treat analysis in a randomized trial, an association between such a genetic variant and the outcome implies a causal effect of the risk factor \citep{vanderweele2014}. Additionally, under further parametric assumptions, the magnitude of the causal effect of the risk factor on the outcome can be estimated \citep{didelez2010}. In this paper, we assume that the effect of the risk factor on the outcome is linear with no effect modification, and the associations of the genetic variants with the risk factor and with the outcome are linear without effect modification \citep{didelez2007}:
\begin{align}
\mathbb{E}(X | G_j=g, U=u)         &= \beta_{X0j} + \beta_{Xj} \; g + \beta_{XU} \; u \label{assumpts} \\
\mathbb{E}(Y | G_j=g, U=u)         &= \beta_{Y0j} + \beta_{Yj} \; g + \beta_{YU} \; u \quad \mbox{ for } j = 1, \ldots, J \notag \\
\mathbb{E}(Y | \dopearl(X=x), U=u) &= \beta_{0}   + \beta \; x      + \beta_{U}  \; u \notag
\end{align}
where $X$ is the risk factor, $G_1, \ldots, G_J$ are the genetic variants, $Y$ is the outcome, $U$ is an unmeasured confounder, $\dopearl(X=x)$ is the do-operator of Pearl meaning that the value of the risk factor is set to $x$ by intervention \citep{pearl2000}, and the causal effect parameter $\beta = \frac{\beta_{Yj}}{\beta_{Xj}}$ for all $j = 1, \ldots, J$. We also assume that the effects of the genetic variants on the risk factor are the same in all individuals. Although these assumptions are not necessary to identify a causal parameter (weaker assumptions have been proposed \citep{swanson2013}), alternative assumptions mean that the causal parameters identified by different instrumental variables are likely to be different. While these assumptions are restrictive, a causal estimate has an interpretation as a test statistic for the null hypothesis that the risk factor is not causal for the outcome without requiring the assumptions of linearity and homogeneity of the genetic effects on the risk factor \citep{burgess2015beyond}.

We assume that summarized data are available in the form of association estimates (beta-coefficients and standard errors) with the risk factor and with the outcome for $j = 1, \ldots, J$ genetic variants that are instrumental variables. The association estimates with the risk factor are denoted $\hat{\beta}_{Xj}$ with standard error $\sigma_{Xj}$; association estimates with the outcome are denoted $\hat{\beta}_{Yj}$ with standard error $\sigma_{Yj}$. The genetic variants are assumed to be independently distributed (that is, not in linkage disequilibrium).

\subsection{Standard inverse-variance weighted method}
The ratio estimate of the causal effect of the risk factor on the outcome based on the $j$th genetic variant is $\frac{\hat{\beta}_{Yj}}{\hat{\beta}_{Xj}}$ \citep{lawlor2007}. We refer to this as $\hat{\beta}_{IVj}$. The variance of the ratio of two random variables can be calculated using the delta method; the formula including first- and second-order terms for the variance of $\hat{\beta}_{IVj}$ is:
\begin{equation}
\var(\hat{\beta}_{IVj}) = \frac{\sigma_{Yj}^2}{\hat{\beta}_{Xj}^2} + \frac{\hat{\beta}_{Yj}^2 \sigma_{Xj}^2}{\hat{\beta}_{Xj}^4} - \frac{2 \theta \hat{\beta}_{Yj} \sigma_{Yj} \sigma_{Xj}}{\hat{\beta}_{Xj}^3} \label{secondvarwith}
\end{equation}
where $\theta$ is the correlation between $\hat{\beta}_{Yj}$ and $\hat{\beta}_{Xj}$ \citep{thomas2007}. This can be rewritten in terms of the causal estimates $\hat{\beta}_{IVj}$ as:
\begin{equation}
\var(\hat{\beta}_{IVj}) = \frac{1}{\hat{\beta}_{Xj}^2} \left(\sigma_{Yj}^2 + \hat{\beta}_{IVj}^2 \sigma_{Xj}^2 - 2 \theta \hat{\beta}_{IVj} \sigma_{Yj} \sigma_{Xj}\right) \label{secondvarwith.2}
\end{equation}

Assuming that the correlation between $\hat{\beta}_{Yj}$ and $\hat{\beta}_{Xj}$ is zero (this would be the case if the associations with the risk factor and with the outcome were estimated in non-overlapping datasets -- known as a two-sample analysis \citep{pierce2013}), the variance is:
\begin{equation}
\var(\hat{\beta}_{IVj}) = \frac{\sigma_{Yj}^2}{\hat{\beta}_{Xj}^2} + \frac{\hat{\beta}_{Yj}^2 \sigma_{Xj}^2}{\hat{\beta}_{Xj}^4} \label{secondvar}.
\end{equation}

If only the first-order term from the delta formula is taken, then the variance is:
\begin{equation}
\var(\hat{\beta}_{IVj}) = \frac{\sigma_{Yj}^2}{\hat{\beta}_{Xj}^2} \label{firstvar}.
\end{equation}

The inverse-variance weighted (IVW) estimate is a weighted mean of the causal estimates from each genetic variant considered individually:
\begin{equation}
\hat{\beta}_{IVW} = \frac{\sum_j \hat{\beta}_{IVj} \var(\hat{\beta}_{IVj})^{-1}}{\sum_j \var(\hat{\beta}_{IVj})^{-1}}.
\end{equation}
This is equivalent to meta-analysing the causal estimates from each genetic variant using the standard inverse-variance weighted formula (hence the name ``inverse-variance weighted estimate'') under a fixed-effect model \citep{borenstein2009}. Using the first-order variance estimates (equation~\ref{firstvar}), the IVW estimate is:
\begin{equation}
\hat{\beta}_{IVW} = \frac{\sum_j \hat{\beta}_{Yj} \hat{\beta}_{Xj} \sigma_{Yj}^{-2}}{\sum_j \hat{\beta}_{Xj}^2 \sigma_{Yj}^{-2}}.
\end{equation}
This is the same estimate as would be obtained from a weighted linear regression of the $\hat{\beta}_{Yj}$ coefficients on the $\hat{\beta}_{Xj}$ coefficients with no intercept term, using the $\sigma_{Yj}^{-2}$ as weights.

Using the first-order weights and assuming a fixed-effect model (Section~\ref{sec:fixed}), the standard error is:
\begin{equation}
\se(\hat{\beta}_{IVW}) = \sqrt{\frac{1}{\sum_j \hat{\beta}_{Xj}^2 \sigma_{Yj}^{-2}}}.
\end{equation}
This is the form of the inverse-variance weighted estimate as it was initially proposed \citepalias{johnson2013vig, ehret2011, dastani2012}.

\subsection{Equivalence to two-stage least squares estimate}
The inverse-variance weighted estimate using first-order weights is also equal to the estimate obtained from the two-stage least squares method that is commonly used with individual-level data (sample size $N$). If the we write the risk factor as $X$ (usually an $N \times 1$ matrix, although the result can be generalized for multiple risk factors \citep{burgess2014pleioajeappendix}), the outcome as $Y$ (an $N \times 1$ matrix), and the instrumental variables as $Z$ (an $N \times J$ matrix), then the two-stage least squares estimate of causal effects \citep{baum2003} is:
\begin{equation}
\hat{\beta}_{2SLS} = [X^T Z (Z^T Z)^{-1} Z^T X]^{-1} X^T Z (Z^T Z)^{-1} Z^T Y. \notag
\end{equation}
This estimate can be obtained by sequential regression of the risk factor on the instrumental variables, and then the outcome on fitted values of the risk factor from the first-stage regression.

Regression of $Y$ on $Z$ gives beta-coefficients $\hat{\beta}_Y = (Z^T Z)^{-1} Z^T Y$ with standard errors the square roots of the diagonal elements of the matrix $(Z^T Z)^{-1} \sigma^2$ where $\sigma$ is the residual standard error. If the instrumental variables are perfectly uncorrelated, then the off-diagonal elements of $(Z^T Z)^{-1} \sigma^2$ are all equal to zero. Regression of $X$ on $Z$ gives beta-coefficients $\hat{\beta}_X = (Z^T Z)^{-1} Z^T X$. Weighted linear regression of the beta-coefficients $\hat{\beta}_Y$ on the beta-coefficients $\hat{\beta}_X$ using the inverse-variance weights $(Z^T Z) \sigma^{-2}$ gives an estimate:
\begin{align}
        \hphantom{=}& [\hat{\beta}_X^T (Z^T Z) \hat{\beta}_X]^{-1} \sigma^{-2} \hat{\beta}_X^T (Z^T Z) \sigma^2 \hat{\beta}_Y \notag \\
                  =& [X^T Z (Z^T Z)^{-1} (Z^T Z) (Z^T Z)^{-1} Z^T X]^{-1} X^T Z (Z^T Z)^{-1} (Z^T Z) (Z^T Z)^{-1} Z^T Y \notag \\
                  =& [X^T Z (Z^T Z)^{-1} Z^T X]^{-1} X^T Z (Z^T Z)^{-1} Z^T Y \notag \\ 
                  =& \hat{\beta}_{2SLS} \notag
\end{align}
The assumption of uncorrelated instrumental variables ensures that the regression coefficients from univariate regressions (as in the regression-based methods) equal those from multivariable regression (as in the two-stage least squares method). In practice, the two-stage least squares and weighted regression-based estimates will differ slightly as there will be non-zero correlations between the genetic variants in finite samples, even if the variants are truly uncorrelated in the population. However, these differences are likely to be slight, and to tend to zero asymptotically \citep{burgess2015scoretj}.

\subsection{Fixed- versus random-effects}
\label{sec:fixed}
A fixed-effect meta-analysis assumes that the causal effects targeted by each genetic variant are all equal. While this would be true if all the genetic variants are valid instrumental variables, and also under the additional linearity assumptions stated above, this may not be true in practice. For instance, genetic variants may affect the exposure via different mechanisms, leading to different magnitudes of effect on the outcome. Alternatively, some variants may have direct effects on the outcome that do not pass via the risk factor, and hence not all genetic variants may be valid instrumental variables. To combat heterogeneity in the causal effects identified by each genetic variant, a random-effects meta-analysis may be preferred. We outline two ways  to model this heterogeneity: an additive random-effects model, and a multiplicative random-effects model.

\subsection{Additive and multiplicative random-effects models}
In a fixed-effect meta-analysis, we assume that the estimates from each instrumental variable $\hat{\beta}_{IVj}$ can be modelled as normally distributed with common mean $\beta_j = \beta$ and variance $\sigma_{IVj}^2$. In a random-effects meta-analysis, the mean values $\beta_j$ are additionally assumed to vary \citep{higgins2009}. In an additive random-effects model, the $\beta_j$ are assumed to be normally distributed with mean $\mu_{\beta}$ and variance $\phi_A^2$. Any additional variability beyond that predicted by the fixed-effect model ($\phi_A > 0$) is interpreted as heterogeneity between the causal effects targeted by each instrumental variable. An estimate of the heterogeneity parameter $\hat{\phi}_A$ is often obtained by a method of moments estimator, developed by DerSimonian and Laird \citep{dersimonian1986}.

In a multiplicative random-effects model, the $\hat{\beta}_{IVj}$ estimates are assumed to be normally distributed with mean $\beta$ and variance $\phi_M^2 \sigma_{IVj}^2$. This model can be fitted by linear regression of the $\hat{\beta}_{Yj}$ on the $\hat{\beta}_{Xj}$ using the $\sigma_{Yj}^{-2}$ as weights. A fixed-effect model can be fitted by setting the residual standard error in the regression model to be one; this can be achieved after fitting the regression model by dividing the standard error by the estimate of the residual standard error \citep{thompson1999}. A multiplicative random-effects model can be fitted by allowing the residual standard error (which is equivalent to the heterogeneity parameter $\phi_M$) to be estimated as part of the model. The multiplicative random-effects model is therefore equivalent to an overdispersed regression model. In case of underdispersion (that is, the estimated residual standard error is less than one), the standard errors should be fixed by setting $\hat{\phi}_M = 1$, as any underdispersion is assumed to occur by chance, and not to be empirically justified.

\begin{align}
\hat{\beta}_{IVj} \sim \mathcal{N}(\beta, \sigma_{IVj}^2) \hphantom{\Big \}} &\qquad\mbox{(fixed-effect model)} \notag \\
\begin{array}{lr}\hat{\beta}_{IVj} &\sim \mathcal{N}(\beta_j, \sigma_{IVj}^2) \\
                 \beta_j           &\sim \mathcal{N}(\beta,   \phi_A^2) \end{array} \Big \}
                                                           &\qquad\mbox{(additive random-effects model)} \notag \\
\hat{\beta}_{IVj} \sim \mathcal{N}(\beta, \phi_M^2 \sigma_{IVj}^2) \hphantom{\Big \}} &\qquad\mbox{(multiplicative random-effects model)} \notag
\end{align}

The point estimate from a fixed-effect meta-analysis is identical to that from a multiplicative random-effects meta-analysis \citep{thompson1999}. However, it differs to that from an additive random-effects meta-analysis when $\hat{\phi}_A \neq 0$, as the weights in the random-effects meta-analysis are inflated to account for heterogeneity. As heterogeneity increases, weights become more similar, which results in estimates with low weights being upweighted (relatively speaking) in an additive random-effects meta-analysis.

\subsection{Weak instrument bias}
Although instrumental variable estimates are consistent (and so they are asymptotically unbiased), they can suffer from substantial bias in finite samples \citep{staiger1997, stock2002}. This bias, known as `weak instrument bias', occurs when the instrumental variables explain a small proportion of variance in the risk factor \citep{burgess2010weak}. In a conventional Mendelian randomization analysis in which the risk factor and outcome are measured in the same participants (a one-sample analysis), weak instrument bias is in the direction of the observational association between the risk factor and the outcome \citep{burgess2010avoiding}. It can also lead to overly narrow confidence intervals and overrejection of the causal null hypothesis \citep{stock2005}. Bias from the inverse-variance weighted method using the first-order weights and a fixed-effect model has been shown to be similar to that from the two-stage least squares method in a realistic simulation study \citep{burgess2013genepi}. However, bias and coverage properties have not been investigated for different choices of the weights or for random-effects models.

\section{Motivating example: analysis of the causal effect of early menopause on triglycerides}
\label{sec:motivate}
This paper was motivated by a particular implementation of two versions of the inverse-variance weighted method with different choices of weights that gave substantially different answers. A Mendelian randomization analysis was performed to assess the causal effect of early menopause risk on triglycerides using 47 genetic variants. Associations of the genetic variants with early menopause (and their standard errors) were obtained from \cite{day2015}; associations represent number of years earlier menopause per additional effect allele. Associations of the genetic variants with triglycerides (and their standard errors) were obtained from the \cite{willer2013}. These associations are provided in Appendix Table~\ref{summdata1} and displayed graphically in Appendix Figure~\ref{weakscatter}. Analyses for the motivating example were performed in Microsoft Excel (Windows 2000 version) and R (version 3.1.2) \citep{r312}.

Fixed-effect inverse-variance weighted methods were performed using the second-order weights (equation~\ref{secondvar}) and the first-order weights (equation~\ref{firstvar}). The weights were substantially the same in both cases; 35 out of the 47 weights differed by less than 5\%, and 44 of the weights differed by less than 10\%. Using the second-order weights (equation~\ref{secondvar}), the causal effect of early menopause on triglycerides was estimated as 0.0021 (standard error, 0.0037; 95\% confidence interval: -0.0052, 0.0095). Using the first-order weights (equation~\ref{firstvar}), the causal effect estimate was 0.0103 (standard error, 0.0036; 95\% confidence interval: 0.0032, 0.0175). These estimates represent the change in triglycerides in standard deviation units per 1 year earlier menopause. The applied implications of this analysis are not the focus of this paper, and depend on the validity of the instrumental variable assumptions for the genetic variants used in the analysis. However, the magnitude of the difference between the estimates (over twice the standard error of the estimates) is striking, and the conclusions from the two analyses would be diametrically opposite. In the first case, the causal null hypothesis that early menopause is a causal risk factor for triglycerides would not be rejected ($p = 0.57$), whereas in the second case, the causal null hypothesis would be rejected ($p = 0.005$). By comparison, using the first-order weights and a multiplicative random effects model, the standard error is 0.0103, meaning that the causal null hypothesis would not be rejected ($p = 0.32$).

\begin{figure}[htbp]
\begin{center}
\includegraphics[width=7cm]{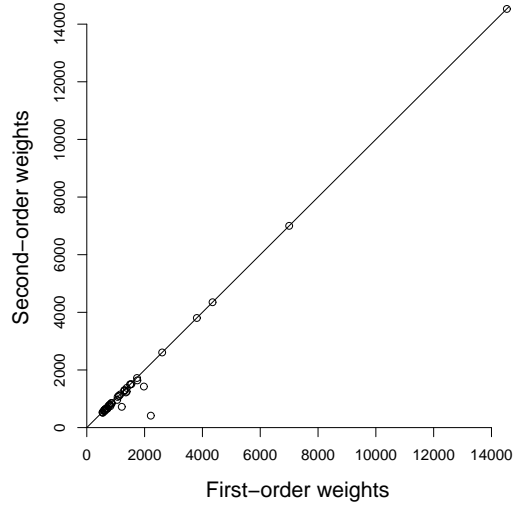}
\end{center}
\caption{Graph of second-order weights against first-order weights for motivating example.} \label{weightsdiff}.
\end{figure}

It turns out that the genetic variant with the greatest difference between the first- and second-order weights is rs704795, the variant that also has the greatest causal estimate. The estimate from this variant is heavily downweighted in the analysis using the second-order weights compared with using the first-order weights. Omitting this variant from the analysis led to similar estimates using the second- and first-order weights (0.0000 versus $-0.0001$). Another interesting observation is that use of the second-order weights reduced heterogeneity between the causal estimates from each genetic variant (for example, in the multiplicative random-effects model, $\hat{\phi}_M$ was 1.69 using the second-order weights compared with 2.83 using the first-order weights). This suggests that, even though the second-order standard errors for the causal estimates from the individual variants will always be greater than the first-order standard errors, precision of the overall causal estimate under a random-effects model may be improved by using the second-order weights when there is heterogeneity between the causal estimates (in this example, $\se(\hat{\beta}_{IVW}) = 0.0063$ in the multiplicative random-effects model using the second-order weights, $\se(\hat{\beta}_{IVW}) = 0.0103$ using the first-order weights).

Estimates from each of the methods are summarized in Table~\ref{motivresults}.

\begin{table}[hbtp]
\begin{minipage}{\textwidth}
\begin{adjustwidth}{-0.5cm}{-0.5cm}
\begin{footnotesize}
\begin{center}
\centering
\begin{tabular}[c]{p{0.3\textwidth}cccc}
\hline
\centering Method                           & Estimate  & Standard error & 95\% confidence interval & Heterogeneity ($\hat{\phi}$) \\
\hline
Fixed-effects model, second-order weights                  &  0.0021   &  0.0037        &  -0.0052,  0.0095        &  -                           \\ [1mm]
Fixed-effects model, first-order weights                   &  0.0103   &  0.0036        &   0.0032,  0.0175        &  -                           \\ [1mm]
Additive random-effects model, second-order weights        &  0.0057   &  0.0070        &  -0.0080,  0.0193        &  0.036                       \\ [1mm]
Additive random-effects model, first-order weights         &  0.0104   &  0.0109        &  -0.0110,  0.0317        &  0.068                       \\ [1mm]
Multiplicative random-effects model, second-order weights  &  0.0021   &  0.0063        &  -0.0103,  0.0145        &  1.686                       \\ [1mm]
Multiplicative random-effects model, first-order weights   &  0.0103   &  0.0103        &  -0.0099,  0.0305        &  2.826                       \\
\hline
\end{tabular}
\caption{Results from motivating example: estimates of causal effect of early menopause on triglycerides (standard deviation increase in triglycerides per one year earlier menopause) using different inverse-variance weighted methods.} \label{motivresults}
\end{center}
\end{footnotesize} 
\end{adjustwidth}
\end{minipage}
\end{table}

In general, genetic variants that have large values of $\hat{\beta}_{Yj}$ and/or small values of $\hat{\beta}_{Xj}$ will be downweighted by the second-order weights. This means that genetic variants that have large and heterogeneous effects on the outcome compared with other variants and/or are weak will be downweighted. Further methodological investigation is therefore needed to investigate the impact on the bias and coverage properties of inverse-variance weighted methods for Mendelian randomization analyses, and which of the versions of the method should be preferred in applied practice.

\section{Simulation study}
\label{sec:simulate}
In this manuscript, we consider estimates from the inverse-variance weighted method using weights from equations (\ref{secondvar}, second-order) and (\ref{firstvar}, first-order), and fixed-effect, additive random-effects, and multiplicative random-effects models for combining the estimates from different genetic variants. Code for implementing these methods is provided in the Appendix. Analyses for the simulation study were performed in R (version 3.1.2).

The data-generating model is as follows:
\begin{align}
z_{ij}   &\sim \mbox{Binomial}(2, 1/3) \mbox{ independently for $j = 1, \ldots, 20$}  \hphantom{Extrat}    \label{genersim}   \\
x_i      &= \sum_{j=1}^{20} \alpha_j z_{ij} + u_i + \epsilon_{Xi} \notag     \\
y_i      &= \beta_X x_i + \sum_{j=1}^{20} \beta_{Zj} z_{ij}  + \beta_U u_i + \epsilon_{Yi} \notag \\
u_i, \epsilon_{Xi}, \epsilon_{Yi} &\sim \mathcal{N}(0, 1) \mbox{ independently} \notag \\
\alpha_j &\sim \mathcal{N}(\alpha, 0.02^2) \mbox{ independently}. \notag
\end{align}
Individuals are indexed by $i$. The 20 genetic variants $z_{ij}$, indexed by $j$, are drawn from binomial distributions, corresponding to single nucleotide polymorphisms (SNPs) with minor allele frequency $1/3$. The risk factor $x_i$ is a linear combination of the genetic variants, a confounder ($u_i$) that is assumed to be unmeasured, and an independent error term ($\epsilon_{Xi}$). The risk factor $y_i$ is a linear combination of the genetic variants, the risk factor, confounder, and a further independent error term ($\epsilon_{Yi}$). The per allele effects of the genetic variants on the risk factor ($\alpha_j$) are drawn from a normal distribution with mean $\alpha$ and variance $0.02^2$. The direct effects of the genetic variants on the outcome ($\beta_{Zj}$, these effects are not via the risk factor) are zero when the genetic variants are valid instrumental variables. The causal effect of the risk factor on the outcome, the main parameter of interest, is $\beta_X$. The effect of the confounder on the outcome is $\beta_U$.

We consider four scenarios:
\begin{enumerate}
\item a one-sample analysis in which the genetic variants are all valid instrumental variables;
\item a one-sample analysis in which the genetic variants have direct effects on the outcome;
\item a two-sample analysis in which the genetic variants are all valid instrumental variables;
\item and a two-sample analysis in which the genetic variants have direct effects on the outcome.
\end{enumerate}

In scenarios 1 and 2, data are generated for $N = 5000$ participants, and associations with the risk factor and with the outcome are estimated in these participants. In scenarios 3 and 4, data are generated for $N = 10\thinspace000$ participants. Associations with the risk factor are estimated in the first 5000 participants, and associations with the outcome in the second 5000 participants. Two-sample analyses are common in Mendelian randomization, particularly when the association estimates are obtained from publicly available data sources \citep{burgess2014twosample}. In a two-sample analysis, weak instrument bias acts in the direction of the null, and hence should not lead to misleading inferences \citep{pierce2013}. However, it is common that many participants in large genetic consortia overlap, such that even if the associations with the risk factor and with the outcome are obtained from separate consortia, they may not be estimated in separate participants. Hence, the one-sample and two-sample settings are both of interest in this paper.

In scenarios 1 and 3, the $\beta_{Zj}$ parameters are all set to zero, and the genetic variants are all valid instrumental variables. In scenarios 2 and 4, the $\beta_{Zj}$ parameters are drawn from a normal distribution with mean 0 and variance $0.02^2$. This is a situation known as ``balanced pleiotropy'' \citep{bowden2015median}. Pleiotropy refers to a genetic variant having an independent effect on the outcome that is not via the risk factor \citep{daveysmith2014}. Balanced pleiotropy means that the pleiotropic effects for all strengths of instrument have mean zero. Such pleiotropic effects should induce heterogeneity between the causal estimates using different genetic variants. Simulations conducted under a multiplicative random-effects model with balanced pleiotropy have suggested that estimates may not be biased on average \citep{bowden2015}. Additional simulations for the case of directional (that is, unbalanced) pleiotropy are considered in the Appendix.

Four sets of parameters are considered -- two values of the causal effect: $\beta_X = 0$ (null causal effect), and $\beta_X = 0.2$ (positive causal effect); and two values of the confounder effect: $\beta_U = +1$ (positive confounding), and $\beta_U = -1$ (negative confounding). Additionally, four values of instrument strength are considered for each set of parameters: $\alpha = 0.03, 0.05, 0.08, 0.10$. 10\thinspace000 simulated datasets are generated in each case.

\subsection{Results}
\noindent \noindent \textbf{Scenarios 1 and 2:} Results from scenario 1 (one-sample, valid instruments) and scenario 2 (one-sample, invalid instruments) are presented in Table~\ref{results.1}. For each value of the instrument strength, set of parameters, and scenario, the mean estimate and empirical power of the 95\% confidence interval (estimate plus or minus 1.96 times the standard error) to reject the null hypothesis is given. The coverage is 100\% minus the power; power under the null hypothesis should be 5\%. The Monte Carlo standard error for the mean estimate is around 0.001 or less, and for the power is 0.2\% under the null, and at most 0.5\% otherwise. Additionally, to judge the instrument strength, the mean F statistic and the mean coefficient of determination ($R^2$ statistic) are given in each case.

With a null causal effect, the results demonstrate the well-known bias and inflated Type 1 error rate of instrumental variable estimates with weak instruments in a one-sample setting. Although bias is similar for both choices of weights (slightly less with the first-order weights), coverage rates are much worse with the first-order weights. Neither the additive nor the multiplicative random-effects models detect heterogeneity in the vast majority of cases (particularly for weaker instruments) with the second-order weights. With the first-order weights, heterogeneity is detected in a greater proportion of simulated datasets. In scenario 1, heterogeneity is not present in the underlying data-generating model, and only estimated by chance; in scenario 2, heterogeneity is expected. For the second-order weights, coverage properties are similar in scenarios 1 and 2; whereas for the first-order weights, coverage properties are worse in scenario 2 for the fixed-effect model, but improved for the random-effects models. For weaker instruments, coverage properties are best using the second-order weights, whereas for stronger instruments, estimates using the first-order weights and a random-effects model perform almost as well, and occasionally better particularly when there is heterogeneity (scenario 2). However, there is inflation of Type 1 error rates even in the best-case scenarios.

With a positive causal effect, estimates with the first-order weights generally have better power to detect a causal effect than those using the second-order weights, particularly with weaker instruments. However, in the light of the Type 1 error rate inflation, this property should not be overvalued. Making fewer Type 2 errors (fewer false negative findings) at the expense of making more Type 1 errors (more false positive findings) is not generally a desirable trade-off.

Additional results from scenarios 1 and 2 are presented in Appendix Table~\ref{results.1a}. For each value of the instrument strength, the (Monte Carlo) standard deviation and the mean standard error of estimates are presented. This helps judge whether uncertainty in the effect estimates is correctly accounted for in the standard errors.

The estimates using second-order weights are the least variable throughout, with the lowest standard deviations. The standard deviation of estimates using second-order weights was always less than the mean standard error of the estimates. In contrast, for scenario 1, the estimates using first-order weights were more variable, but generally had lower average standard errors. This was always true for the fixed-effect analyses, and usually true for the random-effects analyses. However, when there was heterogeneity in the causal estimates identified by the instrumental variables (scenario 2), mean standard errors for the random-effects analyses using first-order weights could be greater than those using second-order weights, despite the second-order standard errors for each causal estimate being uniformly than the first-order standard errors. In scenario 2, mean standard errors for the fixed-effect analyses were generally similar to those in scenario 1, but the standard deviations of the estimates were increased. For the random-effects analyses using the first-order weights in scenario 2, mean standard errors and standard deviations were similar in magnitude. However, mean standard errors using the second-order weights were typically slightly lower, with no loss in coverage (recall Table~\ref{results.1}).

Under the null, standard deviations and mean standard errors are similar whether there is positive or negative confounding, whereas under the alternative, standard errors appear to be wider when confounding is in the same direction as the causal effect, and narrower when confounding is in the opposite direction. This has previously been observed \citep{burgess2011correl}; see Figure 3 of that reference for a potential explanation.

\setlength{\tabcolsep}{4pt}
\begin{table}[hbtp]
\begin{minipage}{\textwidth}
\begin{adjustwidth}{-0.5cm}{-0.5cm}
\begin{footnotesize}
\begin{center}
\centering
\begin{tabular}[c]{ccc|cc|cc|cc|cc|cc|cc}
\hline
\multicolumn{3}{c|}{}         & \multicolumn{4}{c|}{Fixed-effect}  & \multicolumn{4}{c|}{Additive random-effects} & \multicolumn{4}{c}{Multiplicative random-effects} \\
\cline{4-15}
\multicolumn{3}{c|}{}         & \multicolumn{2}{c|}{Second-order}  &  \multicolumn{2}{c|}{First-order} &
                                \multicolumn{2}{c|}{Second-order}  &  \multicolumn{2}{c|}{First-order} &
                                \multicolumn{2}{c|}{Second-order}  &  \multicolumn{2}{c}{First-order}  \\
\hline
$\alpha$ &  $F$     &  $R^2$  &  Mean  &  Power  &  Mean  &  Power  &  Mean  &  Power  &  Mean  &  Power  &  Mean  &  Power  &  Mean  &  Power  \\
\hline
\multicolumn{15}{c}{Scenario 1. One-sample setting, valid instrumental variables}                                                               \\
\hline
\multicolumn{15}{c}{1a. Null causal effect ($\beta_X = 0$), positive confounding ($\beta_U = +1$).}                                             \\
\hline
0.03     &  2.4     & 1.0\%   &  0.166 &  11.4   &  0.200 &  28.5   &  0.166 &  11.4   &  0.196 &  24.3   &  0.166 &  11.4   &  0.200 &  26.2   \\
0.05     &  4.2     & 1.7\%   &  0.111 &   9.9   &  0.118 &  18.3   &  0.111 &   9.9   &  0.110 &  14.6   &  0.111 &   9.9   &  0.118 &  16.3   \\
0.08     &  8.6     & 3.3\%   &  0.058 &   8.1   &  0.052 &  11.1   &  0.057 &   8.0   &  0.044 &   8.5   &  0.058 &   8.0   &  0.052 &   9.6   \\
0.10     & 12.5     & 4.8\%   &  0.041 &   7.2   &  0.036 &   9.4   &  0.040 &   7.0   &  0.029 &   7.3   &  0.041 &   7.2   &  0.036 &   8.1   \\
\hline
\multicolumn{15}{c}{1b. Null causal effect ($\beta_X = 0$), negative confounding ($\beta_U = -1$).}                                             \\
\hline
0.03     &  2.4     & 1.0\%   & -0.170 &  11.2   & -0.207 &  29.1   & -0.170 &  11.2   & -0.204 &  25.2   & -0.170 &  11.2   & -0.207 &  27.1   \\
0.05     &  4.2     & 1.7\%   & -0.108 &   9.3   & -0.113 &  17.7   & -0.107 &   9.3   & -0.105 &  14.3   & -0.108 &   9.3   & -0.113 &  16.0   \\
0.08     &  8.6     & 3.3\%   & -0.059 &   8.2   & -0.054 &  11.2   & -0.059 &   8.1   & -0.046 &   8.5   & -0.059 &   8.2   & -0.054 &   9.7   \\
0.10     & 12.5     & 4.8\%   & -0.041 &   7.1   & -0.036 &   9.4   & -0.040 &   6.8   & -0.029 &   7.2   & -0.041 &   6.9   & -0.036 &   7.8   \\
\hline
\multicolumn{15}{c}{1c. Positive causal effect ($\beta_X = +0.2$), positive confounding ($\beta_U = +1$).}                                      \\
\hline
0.03     &  2.4     & 1.0\%   &  0.337 &  44.8   &  0.403 &  71.9   &  0.337 &  44.8   &  0.402 &  70.1   &  0.337 &  44.8   &  0.403 &  71.2   \\
0.05     &  4.2     & 1.7\%   &  0.284 &  58.3   &  0.318 &  74.1   &  0.284 &  58.3   &  0.315 &  71.5   &  0.284 &  58.3   &  0.318 &  73.2   \\
0.08     &  8.6     & 3.3\%   &  0.241 &  79.5   &  0.257 &  85.4   &  0.241 &  79.4   &  0.254 &  82.5   &  0.241 &  79.5   &  0.257 &  84.5   \\
0.10     & 12.5     & 4.8\%   &  0.222 &  89.6   &  0.234 &  92.4   &  0.222 &  89.5   &  0.231 &  90.5   &  0.222 &  89.6   &  0.234 &  91.7   \\
\hline
\multicolumn{15}{c}{1d. Positive causal effect ($\beta_X = +0.2$), negative confounding ($\beta_U = -1$).}                                      \\
\hline
0.03     &  2.4     & 1.0\%   & -0.006 &   0.8   &  0.000 &   5.1   & -0.006 &   0.8   &  0.006 &   3.4   & -0.006 &   0.8   &  0.000 &   3.8   \\
0.05     &  4.2     & 1.7\%   &  0.061 &   4.3   &  0.092 &  15.6   &  0.061 &   4.3   &  0.106 &  12.7   &  0.061 &   4.3   &  0.092 &  12.2   \\
0.08     &  8.6     & 3.3\%   &  0.114 &  29.7   &  0.146 &  53.1   &  0.114 &  29.7   &  0.160 &  49.6   &  0.114 &  29.6   &  0.146 &  45.7   \\
0.10     & 12.5     & 4.8\%   &  0.132 &  56.4   &  0.160 &  75.6   &  0.134 &  56.2   &  0.172 &  73.2   &  0.132 &  55.5   &  0.160 &  68.1   \\
\hline
\hline
\multicolumn{15}{c}{Scenario 2. One-sample setting, invalid instrumental variables}                                                             \\
\hline
\multicolumn{15}{c}{2a. Null causal effect ($\beta_X = 0$), positive confounding ($\beta_U = +1$).}                                             \\
\hline
0.03     &  2.4     & 1.0\%   &  0.159 &  11.6   &  0.201 &  32.1   &  0.159 &  11.6   &  0.191 &  19.1   &  0.159 &  11.6   &  0.201 &  23.9   \\
0.05     &  4.2     & 1.7\%   &  0.105 &  11.8   &  0.113 &  22.8   &  0.104 &  11.7   &  0.090 &  11.6   &  0.105 &  11.7   &  0.113 &  15.6   \\
0.08     &  8.6     & 3.3\%   &  0.062 &  10.9   &  0.057 &  16.6   &  0.060 &   9.9   &  0.035 &   7.8   &  0.062 &  10.2   &  0.057 &  10.3   \\
0.10     & 12.5     & 4.8\%   &  0.044 &  10.3   &  0.038 &  14.6   &  0.040 &   8.4   &  0.020 &   7.2   &  0.044 &   9.1   &  0.038 &   8.8   \\
\hline
\multicolumn{15}{c}{2b. Null causal effect ($\beta_X = 0$), negative confounding ($\beta_U = -1$).}                                             \\
\hline
0.03     &  2.4     & 1.0\%   & -0.157 &  11.9   & -0.198 &  31.3   & -0.157 &  11.9   & -0.187 &  18.0   & -0.157 &  11.9   & -0.198 &  22.9   \\
0.05     &  4.2     & 1.7\%   & -0.102 &  11.7   & -0.109 &  22.7   & -0.101 &  11.6   & -0.085 &  11.7   & -0.102 &  11.7   & -0.109 &  15.4   \\
0.08     &  8.6     & 3.3\%   & -0.060 &  11.1   & -0.055 &  16.6   & -0.058 &   9.7   & -0.032 &   7.9   & -0.060 &  10.4   & -0.055 &  10.6   \\
0.10     & 12.5     & 4.8\%   & -0.044 &   9.7   & -0.038 &  14.6   & -0.040 &   7.7   & -0.020 &   7.3   & -0.044 &   8.5   & -0.038 &   8.9   \\
\hline
\multicolumn{15}{c}{2c. Positive causal effect ($\beta_X = +0.2$), positive confounding ($\beta_U = +1$).}                                      \\
\hline
0.03     &  2.4     & 1.0\%   &  0.317 &  39.2   &  0.400 &  68.8   &  0.317 &  39.2   &  0.393 &  56.9   &  0.317 &  39.2   &  0.400 &  63.0   \\
0.05     &  4.2     & 1.7\%   &  0.270 &  53.4   &  0.313 &  71.2   &  0.270 &  53.4   &  0.299 &  57.8   &  0.270 &  53.4   &  0.313 &  65.2   \\
0.08     &  8.6     & 3.3\%   &  0.234 &  72.3   &  0.257 &  81.1   &  0.234 &  71.4   &  0.243 &  68.3   &  0.234 &  71.9   &  0.257 &  75.3   \\
0.10     & 12.5     & 4.8\%   &  0.221 &  83.8   &  0.238 &  88.1   &  0.219 &  82.1   &  0.227 &  78.0   &  0.221 &  83.0   &  0.238 &  84.3   \\
\hline
\multicolumn{15}{c}{2d. Positive causal effect ($\beta_X = +0.2$), negative confounding ($\beta_U = -1$).}                                      \\
\hline
0.03     &  2.4     & 1.0\%   & -0.008 &   2.2   & -0.001 &  11.8   & -0.008 &   2.2   &  0.013 &   4.8   & -0.008 &   2.2   & -0.001 &   5.8   \\
0.05     &  4.2     & 1.7\%   &  0.051 &   5.2   &  0.086 &  19.6   &  0.051 &   5.1   &  0.116 &  11.0   &  0.051 &   5.1   &  0.086 &  10.5   \\
0.08     &  8.6     & 3.3\%   &  0.106 &  29.1   &  0.146 &  53.3   &  0.112 &  27.9   &  0.175 &  43.7   &  0.106 &  27.3   &  0.146 &  36.5   \\
0.10     & 12.5     & 4.8\%   &  0.133 &  59.0   &  0.166 &  77.4   &  0.141 &  56.5   &  0.189 &  69.5   &  0.133 &  52.8   &  0.166 &  61.8   \\
\hline
\end{tabular}
\caption{\small{Simulation study results for scenarios 1 and 2 (one-sample setting, valid and invalid instrumental variables): mean estimate and power (\%) of 95\% confidence interval for various inverse-variance weighted methods with four sets of parameter values (null and positive causal effect, positive and negative confounding. The strength of the genetic variants as instruments is indicated: mean per allele effect on the risk factor ($\alpha$), mean F statistic ($F$) and mean coefficient of determination ($R^2$) from regression of risk factor on the genetic variants.}} \label{results.1}
\end{center}
\end{footnotesize} 
\end{adjustwidth}
\end{minipage}
\end{table}
\setlength{\tabcolsep}{6pt}

\vspace{3mm}

\noindent \noindent \textbf{Scenarios 3 and 4:} Results from scenario 3 (two-sample, valid instruments) and scenario 4 (two-sample, invalid instruments) are presented in Table~\ref{results.3} for the mean and power and in Appendix Table~\ref{results.3a} for the standard deviation and standard error. These results demonstrate the well-known bias in the direction of the null in the two-sample setting.

With a null causal effect, no bias is observed. Coverage levels for the second-order weights are conservative, with power below the nominal 5\% level. By contrast, in scenario 3, coverage levels with the first-order weights are close to nominal levels, with slight undercoverage for random-effects models. In scenario 4, there is inflation of Type 1 error rates with the first-order weights for a fixed-effect model, but coverage for both the additive and multiplicative random-effects models is close to nominal levels.

With a positive causal effect, bias is in the direction of the null. The bias is more severe using the second-order weights. Power to detect a causal effect is substantially lower using the second-order weights than using the first-order weights, particularly for weaker instruments.

For the first-order weights, mean standard errors are fairly close to the standard deviations of estimates for the fixed-effect model when there is no heterogeneity in the causal effects, and for the random-effects models when there is heterogeneity in the causal effects. In contrast, for the second-order weights, the mean standard errors are larger than the standard deviations throughout. This corresponds with the coverage properties: in a two-sample setting using first-order weights, estimates are unbiased under the null with correct rejection rates, whereas using second-order weights, rejection rates are conservative.

\vspace{3mm}

\noindent \noindent \textbf{Choice of random-effects model:} As for choosing between the additive and multiplicative random-effects models, with the second-order weights, there was little difference between the results, or even with the results for a fixed-effect model. However, as viewed in the motivating example, there will be a difference if the level of heterogeneity is increased. With the first-order weights, bias was generally slightly less with the additive random-effect model. Coverage under the null was better with an additive random-effects model, and power to detect a causal effect was better with a multiplicative random-effects model. However, differences were slight. Because of the better properties under the null, we therefore prefer the additive random-effects model for the scenarios considered in this paper, although the preference is not a strong one.

\vspace{3mm}

\noindent \noindent \textbf{Directional pleiotropy:} Results with directional pleiotropy are presented in Appendix Table~\ref{results.5}. In brief, the results echo those with no pleiotropy and with balanced pleiotropy: the importance of random-effects models, and the preference for use of second-order weights in a one-sample setting, and first-order weights in a two-sample setting.

\setlength{\tabcolsep}{4pt}
\begin{table}[hbtp]
\begin{minipage}{\textwidth}
\begin{adjustwidth}{-0.5cm}{-0.5cm}
\begin{footnotesize}
\begin{center}
\centering
\begin{tabular}[c]{ccc|cc|cc|cc|cc|cc|cc}
\hline
\multicolumn{3}{c|}{}         & \multicolumn{4}{c|}{Fixed-effect}  & \multicolumn{4}{c|}{Additive random-effects} & \multicolumn{4}{c}{Multiplicative random-effects} \\
\cline{4-15}
\multicolumn{3}{c|}{}         & \multicolumn{2}{c|}{Second-order}  &  \multicolumn{2}{c|}{First-order} &
                                \multicolumn{2}{c|}{Second-order}  &  \multicolumn{2}{c|}{First-order} &
                                \multicolumn{2}{c|}{Second-order}  &  \multicolumn{2}{c}{First-order}  \\
\hline
$\alpha$ &  $F$     &  $R^2$  &  Mean  &  Power  &  Mean  &  Power  &  Mean  &  Power  &  Mean  &  Power  &  Mean  &  Power  &  Mean  &  Power  \\
\hline
\multicolumn{15}{c}{Scenario 3. Two-sample setting, valid instrumental variables}                                                               \\
\hline
\multicolumn{15}{c}{3a. Null causal effect ($\beta_X = 0$), positive confounding ($\beta_U = +1$).}                                             \\
\hline
0.03     &  2.4     & 1.0\%   &  0.001 &   0.8   &   0.002 &  5.1   &   0.001 &  0.8   &   0.002 &  3.7   &   0.001 &  0.8   &   0.002 &  4.2   \\
0.05     &  4.2     & 1.7\%   &  0.001 &   1.2   &   0.001 &  4.8   &   0.001 &  1.2   &   0.001 &  3.6   &   0.001 &  1.2   &   0.001 &  3.9   \\
0.08     &  8.6     & 3.3\%   &  0.001 &   2.0   &   0.001 &  4.9   &   0.001 &  2.0   &   0.001 &  3.8   &   0.001 &  2.0   &   0.001 &  3.9   \\
0.10     & 12.5     & 4.8\%   & -0.001 &   2.1   &  -0.001 &  4.6   &  -0.001 &  2.0   &  -0.001 &  3.6   &  -0.001 &  2.0   &  -0.001 &  3.8   \\
\hline
\multicolumn{15}{c}{3b. Null causal effect ($\beta_X = 0$), negative confounding ($\beta_U = -1$).}                                             \\
\hline
0.03     &  2.4     & 1.0\%   & -0.001 &   0.8   &  -0.001 &  4.9   &  -0.001 &  0.8   &  -0.001 &  3.5   &  -0.001 &  0.8   &  -0.001 &  4.0   \\
0.05     &  4.2     & 1.7\%   &  0.001 &   1.2   &   0.001 &  4.6   &   0.001 &  1.2   &   0.001 &  3.5   &   0.001 &  1.2   &   0.001 &  3.8   \\
0.08     &  8.6     & 3.3\%   & -0.001 &   1.9   &  -0.001 &  4.8   &  -0.001 &  1.9   &  -0.001 &  3.7   &  -0.001 &  1.9   &  -0.001 &  4.0   \\
0.10     & 12.5     & 4.8\%   &  0.000 &   2.4   &   0.000 &  4.8   &   0.000 &  2.4   &   0.000 &  3.8   &   0.000 &  2.4   &   0.000 &  3.9   \\
\hline
\multicolumn{15}{c}{3c. Positive causal effect ($\beta_X = +0.2$), positive confounding ($\beta_U = +1$).}                                      \\
\hline
0.03     &  2.4     & 1.0\%   &  0.090 &   2.4   &  0.119 &  11.5   &  0.090 &   2.4   &  0.121 &   8.8   &  0.090 &   2.4   &  0.119 &   9.7   \\
0.05     &  4.2     & 1.7\%   &  0.126 &  10.2   &  0.162 &  26.1   &  0.126 &  10.2   &  0.166 &  22.0   &  0.126 &  10.2   &  0.162 &  22.6   \\
0.08     &  8.6     & 3.3\%   &  0.153 &  35.5   &  0.182 &  54.1   &  0.153 &  35.2   &  0.186 &  48.8   &  0.153 &  35.3   &  0.182 &  48.9   \\
0.10     & 12.5     & 4.8\%   &  0.162 &  61.8   &  0.184 &  74.7   &  0.162 &  61.2   &  0.188 &  70.7   &  0.162 &  61.1   &  0.184 &  70.5   \\
\hline
\multicolumn{15}{c}{3d. Positive causal effect ($\beta_X = +0.2$), negative confounding ($\beta_U = -1$).}                                      \\
\hline
0.03     &  2.4     & 1.0\%   &  0.091 &   3.7   &  0.120 &  15.7   &  0.091 &   3.7   &  0.123 &  12.2   &  0.091 &   3.7   &  0.120 &  13.1   \\
0.05     &  4.2     & 1.7\%   &  0.121 &  14.4   &  0.154 &  32.9   &  0.121 &  14.4   &  0.159 &  27.6   &  0.121 &  14.4   &  0.154 &  28.7   \\
0.08     &  8.6     & 3.3\%   &  0.152 &  52.0   &  0.181 &  71.6   &  0.152 &  51.8   &  0.185 &  66.8   &  0.152 &  51.7   &  0.181 &  66.8   \\
0.10     & 12.5     & 4.8\%   &  0.159 &  77.6   &  0.182 &  87.8   &  0.160 &  77.1   &  0.185 &  85.0   &  0.159 &  76.9   &  0.182 &  84.5   \\
\hline
\hline
\multicolumn{15}{c}{Scenario 4. Two-sample setting, invalid instrumental variables}                                                             \\
\hline
\multicolumn{15}{c}{4a. Null causal effect ($\beta_X = 0$), positive confounding ($\beta_U = +1$).}                                             \\
\hline
0.03     &  2.4     & 1.0\%   &  0.001 &   2.2   &  0.001 &  10.4   &  0.001 &   2.2   &  0.002 &   4.8   &  0.001 &   2.2   &  0.001 &   6.0   \\
0.05     &  4.2     & 1.7\%   &  0.000 &   2.4   &  0.000 &   9.8   &  0.000 &   2.3   &  0.000 &   4.4   &  0.000 &   2.4   &  0.000 &   5.1   \\
0.08     &  8.6     & 3.3\%   &  0.000 &   4.4   &  0.000 &  10.2   &  0.000 &   4.0   &  0.000 &   5.3   &  0.000 &   4.2   &  0.000 &   5.9   \\
0.10     & 12.5     & 4.8\%   &  0.001 &   5.3   &  0.001 &  10.1   &  0.001 &   4.3   &  0.001 &   5.3   &  0.001 &   4.5   &  0.001 &   5.8   \\
\hline
\multicolumn{15}{c}{4b. Null causal effect ($\beta_X = 0$), negative confounding ($\beta_U = -1$).}                                             \\
\hline
0.03     &  2.4     & 1.0\%   &  0.004 &   1.7   &  0.005 &  10.6   &  0.004 &   1.7   &  0.005 &   4.8   &  0.004 &   1.7   &  0.005 &   6.1   \\
0.05     &  4.2     & 1.7\%   &  0.000 &   3.1   & -0.001 &  10.1   &  0.000 &   3.1   &  0.000 &   4.9   &  0.000 &   3.1   & -0.001 &   5.7   \\
0.08     &  8.6     & 3.3\%   & -0.001 &   4.0   & -0.001 &   9.4   & -0.001 &   3.7   & -0.001 &   5.0   & -0.001 &   3.7   & -0.001 &   5.4   \\
0.10     & 12.5     & 4.8\%   &  0.000 &   5.4   &  0.000 &  10.9   &  0.000 &   4.6   &  0.000 &   6.0   &  0.000 &   4.7   &  0.000 &   6.2   \\
\hline
\multicolumn{15}{c}{4c. Positive causal effect ($\beta_X = +0.2$), positive confounding ($\beta_U = +1$).}                                      \\
\hline
0.03     &  2.4     & 1.0\%   &  0.086 &   3.9   &  0.120 &  16.0   &  0.086 &   3.9   &  0.125 &   8.9   &  0.086 &   3.9   &  0.120 &  10.6   \\
0.05     &  4.2     & 1.7\%   &  0.115 &  10.2   &  0.154 &  27.6   &  0.115 &  10.1   &  0.163 &  17.5   &  0.115 &  10.1   &  0.154 &  19.1   \\
0.08     &  8.6     & 3.3\%   &  0.145 &  34.2   &  0.178 &  54.4   &  0.146 &  33.2   &  0.187 &  42.4   &  0.145 &  33.3   &  0.178 &  42.7   \\
0.10     & 12.5     & 4.8\%   &  0.156 &  56.4   &  0.183 &  71.8   &  0.159 &  53.2   &  0.191 &  61.0   &  0.156 &  52.6   &  0.183 &  60.4   \\
\hline
\multicolumn{15}{c}{4d. Positive causal effect ($\beta_X = +0.2$), negative confounding ($\beta_U = -1$).}                                      \\
\hline
0.03     &  2.4     & 1.0\%   &  0.084 &   5.3   &  0.120 &  21.2   &  0.085 &   5.3   &  0.126 &  10.4   &  0.084 &   5.3   &  0.120 &  12.2   \\
0.05     &  4.2     & 1.7\%   &  0.112 &  14.7   &  0.153 &  36.3   &  0.112 &  14.5   &  0.164 &  21.7   &  0.112 &  14.6   &  0.153 &  23.6   \\
0.08     &  8.6     & 3.3\%   &  0.138 &  46.7   &  0.172 &  66.4   &  0.140 &  44.4   &  0.182 &  52.0   &  0.138 &  44.2   &  0.172 &  52.3   \\
0.10     & 12.5     & 4.8\%   &  0.154 &  70.3   &  0.183 &  83.7   &  0.158 &  66.1   &  0.191 &  72.3   &  0.154 &  65.3   &  0.183 &  71.5   \\
\hline
\end{tabular}
\caption{\small{Simulation study results for scenarios 3 and 4 (two-sample setting, valid and invalid instrumental variables): mean estimate and power (\%) of 95\% confidence interval for various inverse-variance weighted methods with four sets of parameter values (null and positive causal effect, positive and negative confounding. The strength of the genetic variants as instruments is indicated: mean per allele effect on the risk factor ($\alpha$), mean F statistic ($F$) and mean coefficient of determination ($R^2$) from regression of risk factor on the genetic variants.}} \label{results.3}
\end{center}
\end{footnotesize} 
\end{adjustwidth}
\end{minipage}
\end{table}
\setlength{\tabcolsep}{6pt}

\subsection{Additional scenario: extreme outlying variants}
In the motivating example, the difference between estimates seemed to be driven by a single rogue variant. In order to better evaluate bias and coverage in this scenario, we considered an additional simulation scenario 5. Rather than generating the direct effects of the genetic variants on the outcome (the $\beta_{Zj}$ parameters) from a normal distribution with mean 0 and variance $0.02^2$, instead we generated them from a $t$ distribution with 2 degrees of freedom, and multiplied the result by 0.02. The $t$ distribution with a small number of degrees of freedom has much heavier tails than a normal distribution, and so extreme outliers will be more frequent. With 2 degrees of freedom, the variance of the $t$ distribution is not even defined. Simulation results are only considered in the one-sample setting and under the null ($\beta_X = 0$), as inflated Type 1 error rates in this scenario are the primary concern.

In Table~\ref{results.4}, results are given for the inverse-variance weighted methods with different choices of weights and different models for combining the estimates. With a fixed-effect model, coverage rates for the second-order weights are similar to those in scenario 2 with the normally distributed direct effects. For the first-order weights, coverage rates are substantially worse and well above the nominal 5\% level even for the strongest instruments considered in this paper, although bias is similar to that in scenario 2. This corresponds to the motivating example, in which the outlying variant had a large influence on the pooled estimate using the first-order weights, but was heavily downweighted using the second-order weights. However, for a random-effects model using the second-order weights, particularly with the multiplicative random-effects model and for the additive random-effects model with weaker instruments, results were similar to those with a fixed-effect model. In contrast, for a random-effects model with the first-order weights, mean estimates were generally closer to the null (with one notable exception -- scenario 5b, $\alpha = 0.05$ -- that was mostly driven by a single aberrant estimate) and coverage levels were much improved. Coverage levels with a random-effects model were generally slightly better with the first-order weights than with the second-order weights, although not uniformly and the difference was slight. As observed in the motivating example, and particularly with weaker instruments, heterogeneity is more often detected using the first-order weights, as the second-order weights tend to downweight the influence of the outlying variants.

\setlength{\tabcolsep}{4pt}
\begin{table}[hbtp]
\begin{minipage}{\textwidth}
\begin{adjustwidth}{-0.5cm}{-0.5cm}
\begin{footnotesize}
\begin{center}
\centering
\begin{tabular}[c]{ccc|cc|cc|cc|cc|cc|cc}
\hline
\multicolumn{3}{c|}{}         & \multicolumn{4}{c|}{Fixed-effect} & \multicolumn{4}{c|}{Additive random-effects} & \multicolumn{4}{c}{Multiplicative random-effects} \\
\cline{4-15}
\multicolumn{3}{c|}{}         & \multicolumn{2}{c|}{Second-order}  &  \multicolumn{2}{c|}{First-order} &
                                \multicolumn{2}{c|}{Second-order}  &  \multicolumn{2}{c|}{First-order} &
                                \multicolumn{2}{c|}{Second-order}  &  \multicolumn{2}{c}{First-order}  \\
\hline
$\alpha$ &  $F$     &  $R^2$  &  Mean  &  Power  &  Mean  &  Power  &  Mean  &  Power  &  Mean  &  Power  &  Mean  &  Power  &  Mean  &  Power  \\
\hline
\multicolumn{15}{c}{Scenario 5. One-sample setting, invalid instrumental variables with extreme outliers}                                       \\
\hline
\multicolumn{15}{c}{5a. Null causal effect ($\beta_X = 0$), positive confounding ($\beta_U = +1$).}                                             \\
\hline

0.03     &  2.4     & 1.0\%   &  0.149 &  12.5   &  0.200 &  40.7   &  0.149 &  12.3   &  0.208 &  13.2   &  0.149 &  12.4   &  0.200 &  16.6   \\
0.05     &  4.2     & 1.7\%   &  0.105 &  13.1   &  0.106 &  35.3   &  0.103 &  12.2   &  0.078 &   9.0   &  0.105 &  12.5   &  0.106 &  12.4   \\
0.08     &  8.6     & 3.3\%   &  0.058 &  13.9   &  0.053 &  31.9   &  0.050 &   8.9   &  0.005 &   7.4   &  0.058 &  10.2   &  0.053 &   8.9   \\
0.10     & 12.5     & 4.8\%   &  0.042 &  14.0   &  0.035 &  30.6   &  0.030 &   6.9   & -0.004 &   6.8   &  0.042 &   8.1   &  0.035 &   7.6   \\
\hline
\multicolumn{15}{c}{5b. Null causal effect ($\beta_X = 0$), negative confounding ($\beta_U = -1$).}                                             \\
\hline
0.03     &  2.4     & 1.0\%   & -0.153 &  12.8   & -0.209 &  41.4   & -0.152 &  12.6   & -0.178 &  13.7   & -0.153 &  12.7   & -0.209 &  16.6   \\
0.05     &  4.2     & 1.7\%   & -0.101 &  13.8   & -0.094 &  36.0   & -0.100 &  12.4   & -0.232 &   8.6   & -0.101 &  12.9   & -0.094 &  11.5   \\
0.08     &  8.6     & 3.3\%   & -0.058 &  13.4   & -0.067 &  32.6   & -0.050 &   8.4   & -0.005 &   7.0   & -0.058 &  10.0   & -0.067 &   8.9   \\
0.10     & 12.5     & 4.8\%   & -0.043 &  13.9   & -0.047 &  30.2   & -0.032 &   7.2   &  0.001 &   6.4   & -0.043 &   8.6   & -0.047 &   7.9   \\
\hline
\end{tabular}
\caption{\small{Simulation study results for scenarios 5 (one-sample setting, invalid instrumental variables with extreme outliers): mean estimate and power (\%) of 95\% confidence interval for various inverse-variance weighted methods with two sets of parameter values (null causal effect, positive and negative confounding. The strength of the genetic variants as instruments is indicated: mean per allele effect on the risk factor ($\alpha$), mean F statistic ($F$) and mean coefficient of determination ($R^2$) from regression of risk factor on the genetic variants.}} \label{results.4}
\end{center}
\end{footnotesize} 
\end{adjustwidth}
\end{minipage}
\end{table}
\setlength{\tabcolsep}{6pt}

\clearpage

\section{Discussion}
\label{sec:discuss}
Several high-profile Mendelian randomization analyses have employed summarized data and some version of an inverse-variance weighted method. These include analyses of the causal effect of blood pressure on coronary heart disease risk \citepalias{ehret2011}, height on coronary heart disease risk \citep{nelson2015}, adiponectin on type 2 diabetes risk \citep{dastani2012}, lipids on type 2 diabetes risk \citep{fall2015}, and telomere length on risk of various cancers \citep{zhang2015}, amongst several others. The statistical properties of estimates from the inverse-variance weighted method are therefore of considerable interest.

In this paper, we demonstrated that Type 1 error rates for the inverse-variance weighted method as it was initially proposed (first-order weights, fixed-effect model) are likely to be inflated in a one-sample Mendelian randomization setting either when the instruments are weak, or when there is heterogeneity between the causal estimates targeted by different genetic variants. This can be resolved either by using second-order weights or a random-effects model to combine the estimates from multiple genetic variants. These approaches affect the analysis in different ways: the second-order weights tend to downweight the influence of weak and heterogeneous variants on the overall causal estimate, whereas the random-effects models tend to increase standard errors by allowing for heterogeneity between the causal estimates in the model. While both approaches can be applied simultaneously, our simulations indicate that heterogeneity is less substantial when using the second-order weights. However, there is little disadvantage in assuming a random-effects model, as in the absence of heterogeneity, the fixed-effect analysis is recovered, and in the presence of heterogeneity, the random-effects analysis is more appropriate. Our results provide slight preference for an additive random-effects model over a multiplicative random-effects model.

In a two-sample Mendelian randomization setting, weak instruments do not lead to inflated Type 1 error rates but rather attenuate of estimates towards the null. The use of second-order weights was demonstrated to lead to conservative inference, whereas first-order weights gave correct coverage rates under the null. When there was heterogeneity in the causal estimates from different genetic variants, which was simulated to arise due to genetic variants having pleiotropic effects, a fixed-effect model with first-order weights was shown to lead to undercoverage, although this was corrected by use of a random-effects model.

A conclusion from this paper is the need to assess heterogeneity between the causal estimates from different genetic variants prior to performing a Mendelian randomization analysis based on multiple genetic variants, for example by a scatter plot of the gene--risk factor and gene--outcome associations (Appendix Figure~\ref{weakscatter}). The presence of heterogeneous variants is likely to indicate violation of the instrumental variable assumptions for some of the variants, and can lead to misleading estimates as observed in the motivating example. Assessment for heterogeneity is also relevant when performing an analysis using individual-level data, for example using a two-stage least squares or allele score method.

\subsection{Limitation of simulation studies}
Our conclusions are limited as they are based on simulation studies. This is by necessity, as the properties of the estimators that we want to assess are finite-sample properties, not asymptotic properties. Our findings may have differed if we had considered a different data-generating mechanism, or more substantial heterogeneity between estimates from genetic variants. However, the findings are in line with theoretical considerations, and we believe the scenarios that we have chosen to be representative of a typical Mendelian randomization investigation in practice.

\subsection{Unbalanced pleiotropy and robust methods (Egger regression, median-based approaches)}
In particular, we mostly considered scenarios in this paper corresponding to balanced pleiotropy. In the case of unbalanced (or directional) pleiotropy, causal estimates from inverse-variance weighted methods are biased and Type 1 error rates are inflated in all settings, even in the asymptotic limit \citep{bowden2015median}. This can be resolved in a number of ways. In Egger regression, we perform a weighted linear regression of the gene--outcome association estimates ($\hat{\beta}_{Yj}$) on the gene--risk factor association estimates ($\hat{\beta}_{Xj}$) in the same way as in an inverse-variance weighted method, except that an intercept term is included in the regression model. This intercept term represents the average direct effect of the genetic variants on the outcome. (It is additionally required that all genetic variants are orientated such that the $\hat{\beta}_{Xj}$ estimates are all positive, or are all negative.) The causal estimate from Egger regression is the slope parameter from this regression model. It is a consistent estimate of the causal effect under the alternative assumption that the direct effects of the genetic variants are uncorrelated with the instrument strength; this is known as the InSIDE (instrument strength independent of direct effect) assumption. In the notation of the data-generating model of equation~(\ref{genersim}), the $\alpha_j$ parameters must be uncorrelated with the $\beta_{Zj}$ parameters; in the balanced pleiotropy examples of this paper, these parameters are drawn from independent distributions. This is a weaker assumption than the standard instrumental variable assumptions (the $\beta_{Zj}$ parameters all equal zero) or the assumption of balanced pleiotropy (the $\beta_{Zj}$ parameters have mean zero).

Similar considerations as to the choice of weights in Egger regression could be considered; the original proposal was equivalent to using the first-order weights. Informal simulations (not presented) have suggested that the same conclusions from this paper also hold for Egger regression (particularly the use of random-effects models). However, a full investigation would require simulating data with unbalanced pleiotropy (potentially both when the InSIDE assumption is satisfied and when it is violated); this is considered to be beyond the scope of this paper.

One notable difference about Egger regression is that if the genetic variants are allowed to have direct effects on the outcome, then heterogeneity in the causal estimates from individual variants is expected. Therefore, while heterogeneity in an inverse-variance weighted analysis is unwelcome and a potential sign that the assumptions are not satisfied, heterogeneity in the Egger method is a natural consequence of weakening the instrumental variable assumptions and does not necessarily invalidate the analysis.

Another approach for dealing with unbalanced pleiotropy is a median-based approach. The median of the causal estimates from each of the genetic variants taken individually is a consistent estimate of the causal effect under the assumption that at least 50\% of the genetic variants are valid instrumental variables \citep{han2008}. This is a different assumption to the InSIDE assumption, and neither assumption includes all cases of the other. Confidence intervals for the median can be obtained by bootstrapping; we suggest estimating a bootstrap standard error and forming confidence intervals from the standard error \citep{bowden2015median}. A weighted median estimator can also be obtained using inverse-variance weights in a weighted median function \citep{bowden2015median}. This method may have better asymptotic properties than an inverse-variance weighted method in a number of cases, as outlying estimates do not influence the median of the distribution. Simulations performed using second- and first-order weights from the delta method suggested that weighted median estimates were not sensitive to the particular choice of weighting function. In a median-based approach, the choice of weights influences not only the bias and variability of estimates, but also the identification condition, as the consistency criterion for a weighted median estimator is that 50\% of the weight in the analysis corresponds to valid instrumental variables. Hence, in some cases, the simple (unweighted) median estimator may be preferred even if it is less efficient.

\subsection{Overlap between the samples in a `two-sample' analysis}
\label{sec:discuss:overlap}
In practice, before following the recommendation to use first-order weights in a two-sample Mendelian randomization setting, it is advisable to check whether the samples used to estimate the gene--risk factor and the gene--outcome associations truly do not overlap. In the motivating example of the paper, genetic associations with early menopause are obtained from a consortium of 33 studies, and genetic associations with triglycerides from a consortium of 23 studies. Although the consortia appear to be different, in fact, at least 17 of the studies are included in both consortia, meaning that the analysis is not a true two-sample analysis. It is not clear exactly the extent of the overlap without having the individual-level data, but it is likely to be substantial.

Although the full second-order expression for the variance of a causal estimate (equation~\ref{secondvarwith}) includes a term $\theta$ that depends on the overlap between the two datasets, in this paper we have set $\theta = 0$ even in a one-sample setting. This was undertaken for computational simplicity in the simulation study setting. If the individual-level data were available, an estimate of $\theta$ could be obtained by bootstrapping the samples, and calculating the correlation between the bootstrapped distributions of $\hat{\beta}_{Yj}$ and $\hat{\beta}_{Xj}$ for each $j$. However, this was infeasible in the simulation study. Additionally, if the individual-level data are not available, it is unclear how to estimate $\theta$. A sensitivity analysis can be performed for the value of $\theta$; results for the motivating example of this paper are shown in Appendix Table~\ref{results9}. We see that different choices of $\theta$ lead to similar causal estimates and 95\% confidence intervals for each of the inverse-variance weighted methods.

\subsection{Interpretation of a random-effects estimate}
A theoretical concern in recommending the use of random-effects models for Mendelian randomization is the interpretation of the random-effects estimate. Under the assumptions of linearity and no effect modification, and in particular under the stable unit treatment value assumption (SUTVA \citep{cox1958sec2} -- this states that the effect on the outcome of modifying the risk factor should be the same for all possible interventions on the risk factor, also expressed as ``no multiple versions of treatment'' \citep{vanderweele2013}), the causal estimates from different instrumental variables should target the same causal parameter. However, in reality, taking the context of the motivating example, different interventions on age at menopause (such as ooectomy, hysterectomy, and hormone therapy) may have different effects on triglyceride levels; similar heterogeneity is expected for genetic variants that affect age at menopause via different biological pathways. By allowing for heterogeneity in causal estimates from different genetic variants, the notion of a single causal effect of the risk factor on the outcome is lost, and it is not clear for what intervention on the risk factor the causal estimate is targeting. Additionally, if the choice of genetic variants changes, then the causal parameter also changes, as the random-effects distribution is taken across a different set of variants. The random-effects estimate is correctly interpreted not as targeting a common causal effect, but as targeting the average value of the distribution of causal effects identified by the different variants \citep{riley2011}. This subtlety is not unique to causal estimation, rather it is relevant in meta-analysis more widely \citep{higgins2009}. However, heterogeneity is more forgiveable in meta-analysis; it could be argued that any deviation from homogeneity should be interpreted as evidence that the instrumental variable assumptions are violated for at least one of the genetic variants, and so a causal estimate based on all the genetic variants should not be presented.

We take a practical approach, and view these theoretical concerns as secondary to the primary concern of obtaining reliable causal inferences \citep{burgess2015beyond}. Our view is that a literal interpretation of causal effect estimates from Mendelian randomization is rarely justified, due to differences between the way in which genetic variants influence the risk factor and any potential clinical intervention on the risk factor in practice \citep{burgess2012bmj}. However, if there is substantial heterogeneity, or if there are individual genetic variants that clear outliers, then the overall causal estimate is likely to be unreliable even as a test of causality, and the instrumental variable assumptions should be examined carefully, particularly for the outlying variants.

\subsection{Conclusion}
In conclusion, in a Mendelian randomization analysis using summarized data in a (strict) two-sample setting (that is, when there is no overlap between the datasets in which associations with the risk factor and with the outcome are estimated), the inverse-variance weighted method with first-order weights may be preferred, although a random-effects model for combining the causal effects from the individual genetic variants should be used. In a one-sample setting, or if there is any overlap between the datasets, then a random-effects model using the second-order weights should be preferred to avoid false-positive findings. If the overlap is not substantial, then an analysis using the first-order weights may be presented as a sensitivity analysis, as it may have increased power to detect a causal effect.

\bibliographystyle{imsart-nameyear}
\bibliography{masterref}

\begin{thebibliography}{49}

\bibitem[\protect\citeauthoryear{Ahmad et~al.}{2015}]{ahmad2015}
\begin{barticle}[author]
\bauthor{\bsnm{Ahmad},~\bfnm{Omar~S}\binits{O.~S.}},
  \bauthor{\bsnm{Morris},~\bfnm{John~A}\binits{J.~A.}},
  \bauthor{\bsnm{Mujammami},~\bfnm{Muhammad}\binits{M.}},
  \bauthor{\bsnm{Forgetta},~\bfnm{Vincenzo}\binits{V.}},
  \bauthor{\bsnm{Leong},~\bfnm{Aaron}\binits{A.}},
  \bauthor{\bsnm{Li},~\bfnm{Rui}\binits{R.}},
  \bauthor{\bsnm{Turgeon},~\bfnm{Maxime}\binits{M.}},
  \bauthor{\bsnm{Greenwood},~\bfnm{Celia~MT}\binits{C.~M.}},
  \bauthor{\bsnm{Thanassoulis},~\bfnm{George}\binits{G.}},
  \bauthor{\bsnm{Meigs},~\bfnm{James~B}\binits{J.~B.}} \betal{et~al.}
(\byear{2015}).
\btitle{{A Mendelian randomization study of the effect of type-2 diabetes on
  coronary heart disease}}.
\bjournal{Nature Communications}
\bvolume{6}
\bpages{7060}.
\bdoi{10.1038/ncomms8060}
\end{barticle}
\endbibitem

\bibitem[\protect\citeauthoryear{Baum, Schaffer and Stillman}{2003}]{baum2003}
\begin{barticle}[author]
\bauthor{\bsnm{Baum},~\bfnm{CF}\binits{C.}},
  \bauthor{\bsnm{Schaffer},~\bfnm{ME}\binits{M.}} \AND
  \bauthor{\bsnm{Stillman},~\bfnm{S}\binits{S.}}
(\byear{2003}).
\btitle{{Instrumental variables and GMM: Estimation and testing}}.
\bjournal{Stata Journal}
\bvolume{3}
\bpages{1--31}.
\end{barticle}
\endbibitem

\bibitem[\protect\citeauthoryear{Borenstein et~al.}{2009}]{borenstein2009}
\begin{bbook}[author]
\bauthor{\bsnm{Borenstein},~\bfnm{M.}\binits{M.}},
  \bauthor{\bsnm{Hedges},~\bfnm{L.~V.}\binits{L.~V.}},
  \bauthor{\bsnm{Higgins},~\bfnm{J.~P.~T.}\binits{J.~P.~T.}} \AND
  \bauthor{\bsnm{Rothstein},~\bfnm{H.~R.}\binits{H.~R.}}
(\byear{2009}).
\btitle{{Introduction to meta-analysis. Chapter 34: Generality of the basic
  inverse-variance method}}.
\bpublisher{Wiley}.
\end{bbook}
\endbibitem

\bibitem[\protect\citeauthoryear{Bowden, Davey~Smith and
  Burgess}{2015}]{bowden2015}
\begin{barticle}[author]
\bauthor{\bsnm{Bowden},~\bfnm{Jack}\binits{J.}},
  \bauthor{\bsnm{Davey~Smith},~\bfnm{George}\binits{G.}} \AND
  \bauthor{\bsnm{Burgess},~\bfnm{Stephen}\binits{S.}}
(\byear{2015}).
\btitle{{Mendelian randomization with invalid instruments: effect estimation
  and bias detection through Egger regression}}.
\bjournal{International Journal of Epidemiology}
\bvolume{44}
\bpages{512--525}.
\end{barticle}
\endbibitem

\bibitem[\protect\citeauthoryear{Bowden et~al.}{2015}]{bowden2015median}
\begin{bunpublished}[author]
\bauthor{\bsnm{Bowden},~\bfnm{Jack}\binits{J.}},
  \bauthor{\bsnm{Davey~Smith},~\bfnm{George}\binits{G.}},
  \bauthor{\bsnm{Haycock},~\bfnm{Philip~C}\binits{P.~C.}} \AND
  \bauthor{\bsnm{Burgess},~\bfnm{Stephen}\binits{S.}}
(\byear{2015}).
\btitle{{Consistent estimation in Mendelian randomization with some invalid
  instruments using a weighted median estimator}}.
\bnote{Available at https://www.academia.edu/15479132/Consistent}.
\end{bunpublished}
\endbibitem

\bibitem[\protect\citeauthoryear{Burgess, Butterworth and
  Thompson}{2013}]{burgess2013genepi}
\begin{barticle}[author]
\bauthor{\bsnm{Burgess},~\bfnm{S.}\binits{S.}},
  \bauthor{\bsnm{Butterworth},~\bfnm{A.}\binits{A.}} \AND
  \bauthor{\bsnm{Thompson},~\bfnm{S.~G.}\binits{S.~G.}}
(\byear{2013}).
\btitle{{Mendelian randomization analysis with multiple genetic variants using
  summarized data}}.
\bjournal{Genetic Epidemiology}
\bvolume{37}
\bpages{658--665}.
\bdoi{10.1002/gepi.21758}
\end{barticle}
\endbibitem

\bibitem[\protect\citeauthoryear{Burgess, Butterworth and
  Thompson}{2015}]{burgess2015beyond}
\begin{barticle}[author]
\bauthor{\bsnm{Burgess},~\bfnm{Stephen}\binits{S.}},
  \bauthor{\bsnm{Butterworth},~\bfnm{Adam~S}\binits{A.~S.}} \AND
  \bauthor{\bsnm{Thompson},~\bfnm{John~R}\binits{J.~R.}}
(\byear{2015}).
\btitle{{Beyond Mendelian randomization: how to interpret evidence of shared
  genetic predictors}}.
\bjournal{Journal of Clinical Epidemiology}.
\bdoi{10.1016/j.jclinepi.2015.08.001}
\end{barticle}
\endbibitem

\bibitem[\protect\citeauthoryear{Burgess, Dudbridge and
  Thompson}{2015a}]{burgess2014pleioajeappendix}
\begin{barticle}[author]
\bauthor{\bsnm{Burgess},~\bfnm{Stephen}\binits{S.}},
  \bauthor{\bsnm{Dudbridge},~\bfnm{Frank}\binits{F.}} \AND
  \bauthor{\bsnm{Thompson},~\bfnm{Simon~G}\binits{S.~G.}}
(\byear{2015}a).
\btitle{{Re: ``Multivariable Mendelian randomization: the use of pleiotropic
  genetic variants to estimate causal effects''}}.
\bjournal{American Journal of Epidemiology}
\bvolume{181}
\bpages{290--291}.
\end{barticle}
\endbibitem

\bibitem[\protect\citeauthoryear{Burgess, Dudbridge and
  Thompson}{2015b}]{burgess2015scoretj}
\begin{bunpublished}[author]
\bauthor{\bsnm{Burgess},~\bfnm{S}\binits{S.}},
  \bauthor{\bsnm{Dudbridge},~\bfnm{F}\binits{F.}} \AND
  \bauthor{\bsnm{Thompson},~\bfnm{SG}\binits{S.}}
(\byear{2015}b).
\btitle{{Combining information on multiple instrumental variables in Mendelian
  randomization: comparison of allele score and summarized data methods}}.
\bnote{Available at https://www.academia.edu/15479109/Combining}.
\end{bunpublished}
\endbibitem

\bibitem[\protect\citeauthoryear{Burgess and Thompson}{2011}]{burgess2010weak}
\begin{barticle}[author]
\bauthor{\bsnm{Burgess},~\bfnm{S.}\binits{S.}} \AND
  \bauthor{\bsnm{Thompson},~\bfnm{S.~G.}\binits{S.~G.}}
(\byear{2011}).
\btitle{{Bias in causal estimates from Mendelian randomization studies with
  weak instruments}}.
\bjournal{Statistics in Medicine}
\bvolume{30}
\bpages{1312--1323}.
\bdoi{10.1002/sim.4197}
\end{barticle}
\endbibitem

\bibitem[\protect\citeauthoryear{Burgess, Thompson and {CRP CHD Genetics
  Collaboration}}{2011}]{burgess2010avoiding}
\begin{barticle}[author]
\bauthor{\bsnm{Burgess},~\bfnm{S.}\binits{S.}},
  \bauthor{\bsnm{Thompson},~\bfnm{S.~G.}\binits{S.~G.}} \AND
  \bauthor{\bsnm{{CRP CHD Genetics Collaboration}}}
(\byear{2011}).
\btitle{{Avoiding bias from weak instruments in Mendelian randomization
  studies}}.
\bjournal{International Journal of Epidemiology}
\bvolume{40}
\bpages{755--764}.
\bdoi{10.1093/ije/dyr036}
\end{barticle}
\endbibitem

\bibitem[\protect\citeauthoryear{Burgess and
  Thompson}{2012}]{burgess2011correl}
\begin{barticle}[author]
\bauthor{\bsnm{Burgess},~\bfnm{S.}\binits{S.}} \AND
  \bauthor{\bsnm{Thompson},~\bfnm{S.~G.}\binits{S.~G.}}
(\byear{2012}).
\btitle{{Improvement of bias and coverage in instrumental variable analysis
  with weak instruments for continuous and binary outcomes}}.
\bjournal{Statistics in Medicine}
\bvolume{31}
\bpages{1582--1600}.
\bdoi{10.1002/sim.4498}
\end{barticle}
\endbibitem

\bibitem[\protect\citeauthoryear{Burgess and Thompson}{2015}]{burgess2015book}
\begin{bbook}[author]
\bauthor{\bsnm{Burgess},~\bfnm{Stephen}\binits{S.}} \AND
  \bauthor{\bsnm{Thompson},~\bfnm{Simon~G}\binits{S.~G.}}
(\byear{2015}).
\btitle{{Mendelian randomization: methods for using genetic variants in causal
  estimation}}.
\bpublisher{Chapman \& Hall}.
\end{bbook}
\endbibitem

\bibitem[\protect\citeauthoryear{Burgess et~al.}{2012}]{burgess2012bmj}
\begin{barticle}[author]
\bauthor{\bsnm{Burgess},~\bfnm{S}\binits{S.}},
  \bauthor{\bsnm{Butterworth},~\bfnm{A}\binits{A.}},
  \bauthor{\bsnm{Malarstig},~\bfnm{A}\binits{A.}} \AND
  \bauthor{\bsnm{Thompson},~\bfnm{SG}\binits{S.}}
(\byear{2012}).
\btitle{{Use of Mendelian randomisation to assess potential benefit of clinical
  intervention}}.
\bjournal{British Medical Journal}
\bvolume{345}
\bpages{e7325}.
\bdoi{10.1136/bmj.e7325}
\end{barticle}
\endbibitem

\bibitem[\protect\citeauthoryear{Burgess et~al.}{2015}]{burgess2014twosample}
\begin{barticle}[author]
\bauthor{\bsnm{Burgess},~\bfnm{S}\binits{S.}},
  \bauthor{\bsnm{Scott},~\bfnm{RA}\binits{R.}},
  \bauthor{\bsnm{Timpson},~\bfnm{NJ}\binits{N.}},
  \bauthor{\bsnm{Davey~Smith},~\bfnm{G}\binits{G.}},
  \bauthor{\bsnm{Thompson},~\bfnm{SG}\binits{S.}} \AND
  \bauthor{\bsnm{{EPIC-InterAct Consortium}}}
(\byear{2015}).
\btitle{{Using published data in Mendelian randomization: a blueprint for
  efficient identification of causal risk factors}}.
\bjournal{European Journal of Epidemiology}
\bvolume{30}
\bpages{543--552}.
\bdoi{10.1007/s10654-015-0011-z}
\end{barticle}
\endbibitem

\bibitem[\protect\citeauthoryear{{The Global Lipids Genetics
  Consortium}}{2013}]{willer2013}
\begin{barticle}[author]
\bauthor{\bsnm{{The Global Lipids Genetics Consortium}}}
(\byear{2013}).
\btitle{{Discovery and refinement of loci associated with lipid levels}}.
\bjournal{Nature Genetics}
\bvolume{45}
\bpages{1274--1283}.
\bdoi{10.1038/ng.2797}
\end{barticle}
\endbibitem

\bibitem[\protect\citeauthoryear{Cox}{1958}]{cox1958sec2}
\begin{bbook}[author]
\bauthor{\bsnm{Cox},~\bfnm{D.~R.}\binits{D.~R.}}
(\byear{1958}).
\btitle{{Planning of experiments. Section 2: Some key assumptions}}.
\bpublisher{Wiley}.
\end{bbook}
\endbibitem

\bibitem[\protect\citeauthoryear{Dastani et~al.}{2012}]{dastani2012}
\begin{barticle}[author]
\bauthor{\bsnm{Dastani},~\bfnm{Zari}\binits{Z.}},
  \bauthor{\bsnm{Hivert},~\bfnm{Marie-France}\binits{M.-F.}},
  \bauthor{\bsnm{Timpson},~\bfnm{Nicholas}\binits{N.}},
  \bauthor{\bsnm{Perry},~\bfnm{John~RB}\binits{J.~R.}},
  \bauthor{\bsnm{Yuan},~\bfnm{Xin}\binits{X.}},
  \bauthor{\bsnm{Scott},~\bfnm{Robert~A}\binits{R.~A.}},
  \bauthor{\bsnm{Henneman},~\bfnm{Peter}\binits{P.}},
  \bauthor{\bsnm{Heid},~\bfnm{Iris~M}\binits{I.~M.}},
  \bauthor{\bsnm{Kizer},~\bfnm{Jorge~R}\binits{J.~R.}},
  \bauthor{\bsnm{Lyytik{\"a}inen},~\bfnm{Leo-Pekka}\binits{L.-P.}}
  \betal{et~al.}
(\byear{2012}).
\btitle{{Novel loci for adiponectin levels and their influence on type 2
  diabetes and metabolic traits: A multi-ethnic meta-analysis of 45,891
  individuals}}.
\bjournal{PLOS Genetics}
\bvolume{8}
\bpages{e1002607}.
\bdoi{10.1371/journal.pgen.1002607}
\end{barticle}
\endbibitem

\bibitem[\protect\citeauthoryear{Davey~Smith and
  Ebrahim}{2003}]{daveysmith2003}
\begin{barticle}[author]
\bauthor{\bsnm{Davey~Smith},~\bfnm{G}\binits{G.}} \AND
  \bauthor{\bsnm{Ebrahim},~\bfnm{S}\binits{S.}}
(\byear{2003}).
\btitle{{`Mendelian randomization': can genetic epidemiology contribute to
  understanding environmental determinants of disease?}}
\bjournal{International Journal of Epidemiology}
\bvolume{32}
\bpages{1--22}.
\bdoi{10.1093/ije/dyg070}
\end{barticle}
\endbibitem

\bibitem[\protect\citeauthoryear{Davey~Smith and Hemani}{2014}]{daveysmith2014}
\begin{barticle}[author]
\bauthor{\bsnm{Davey~Smith},~\bfnm{George}\binits{G.}} \AND
  \bauthor{\bsnm{Hemani},~\bfnm{Gibran}\binits{G.}}
(\byear{2014}).
\btitle{{Mendelian randomization: genetic anchors for causal inference in
  epidemiological studies}}.
\bjournal{Human Molecular Genetics}
\bvolume{23}
\bpages{R89--98}.
\bdoi{10.1093/hmg/ddu328}
\end{barticle}
\endbibitem

\bibitem[\protect\citeauthoryear{Day et~al.}{2015}]{day2015}
\begin{barticle}[author]
\bauthor{\bsnm{Day},~\bfnm{Felix}\binits{F.}} \betal{et~al.}
(\byear{2015}).
\btitle{{Large-scale genomic analyses link reproductive aging to hypothalamic
  signaling, breast cancer susceptibility and BRCA1-mediated DNA repair}}.
\bjournal{Nature Genetics}.
\bdoi{10.1038/ng.3412}
\end{barticle}
\endbibitem

\bibitem[\protect\citeauthoryear{DerSimonian and Laird}{1986}]{dersimonian1986}
\begin{barticle}[author]
\bauthor{\bsnm{DerSimonian},~\bfnm{R.}\binits{R.}} \AND
  \bauthor{\bsnm{Laird},~\bfnm{N.}\binits{N.}}
(\byear{1986}).
\btitle{{Meta-analysis in clinical trials}}.
\bjournal{Controlled Clinical Trials}
\bvolume{7}
\bpages{177--188}.
\bdoi{10.1016/0197-2456(86)90046-2}
\end{barticle}
\endbibitem

\bibitem[\protect\citeauthoryear{Didelez, Meng and Sheehan}{2010}]{didelez2010}
\begin{barticle}[author]
\bauthor{\bsnm{Didelez},~\bfnm{V.}\binits{V.}},
  \bauthor{\bsnm{Meng},~\bfnm{S.}\binits{S.}} \AND
  \bauthor{\bsnm{Sheehan},~\bfnm{N.~A.}\binits{N.~A.}}
(\byear{2010}).
\btitle{{Assumptions of IV methods for observational epidemiology}}.
\bjournal{Statistical Science}
\bvolume{25}
\bpages{22--40}.
\bdoi{10.1214/09-sts316}
\end{barticle}
\endbibitem

\bibitem[\protect\citeauthoryear{Didelez and Sheehan}{2007}]{didelez2007}
\begin{barticle}[author]
\bauthor{\bsnm{Didelez},~\bfnm{V}\binits{V.}} \AND
  \bauthor{\bsnm{Sheehan},~\bfnm{N}\binits{N.}}
(\byear{2007}).
\btitle{{Mendelian randomization as an instrumental variable approach to causal
  inference}}.
\bjournal{Statistical Methods in Medical Research}
\bvolume{16}
\bpages{309--330}.
\bdoi{10.1177/0962280206077743}
\end{barticle}
\endbibitem

\bibitem[\protect\citeauthoryear{Fall et~al.}{2015}]{fall2015}
\begin{barticle}[author]
\bauthor{\bsnm{Fall},~\bfnm{Tove}\binits{T.}},
  \bauthor{\bsnm{Xie},~\bfnm{Weijia}\binits{W.}},
  \bauthor{\bsnm{Poon},~\bfnm{Wenny}\binits{W.}},
  \bauthor{\bsnm{Yaghootkar},~\bfnm{Hanieh}\binits{H.}},
  \bauthor{\bsnm{M{\"a}gi},~\bfnm{Reedik}\binits{R.}},
  \bauthor{\bsnm{Knowles},~\bfnm{Joshua~W}\binits{J.~W.}},
  \bauthor{\bsnm{Lyssenko},~\bfnm{Valeriya}\binits{V.}},
  \bauthor{\bsnm{Weedon},~\bfnm{Michael}\binits{M.}},
  \bauthor{\bsnm{Frayling},~\bfnm{Timothy~M}\binits{T.~M.}} \AND
  \bauthor{\bsnm{Ingelsson},~\bfnm{Erik}\binits{E.}}
(\byear{2015}).
\btitle{{Using genetic variants to assess the relationship between circulating
  lipids and type 2 diabetes}}.
\bjournal{Diabetes}
\bvolume{doi:10.2337/db14-1710}.
\bdoi{10.2337/db14-1710}
\end{barticle}
\endbibitem

\bibitem[\protect\citeauthoryear{{The International Consortium for Blood
  Pressure Genome-Wide Association Studies}}{2011}]{ehret2011}
\begin{barticle}[author]
\bauthor{\bsnm{{The International Consortium for Blood Pressure Genome-Wide
  Association Studies}}}
(\byear{2011}).
\btitle{{Genetic variants in novel pathways influence blood pressure and
  cardiovascular disease risk}}.
\bjournal{Nature}
\bvolume{478}
\bpages{103--109}.
\bdoi{10.1038/nature10405}
\end{barticle}
\endbibitem

\bibitem[\protect\citeauthoryear{Greenland}{2000}]{greenland2000}
\begin{barticle}[author]
\bauthor{\bsnm{Greenland},~\bfnm{S}\binits{S.}}
(\byear{2000}).
\btitle{{An introduction to instrumental variables for epidemiologists}}.
\bjournal{International Journal of Epidemiology}
\bvolume{29}
\bpages{722--729}.
\bdoi{10.1093/ije/29.4.722}
\end{barticle}
\endbibitem

\bibitem[\protect\citeauthoryear{Han}{2008}]{han2008}
\begin{barticle}[author]
\bauthor{\bsnm{Han},~\bfnm{C.}\binits{C.}}
(\byear{2008}).
\btitle{{Detecting invalid instruments using L1-GMM}}.
\bjournal{Economics Letters}
\bvolume{101}
\bpages{285--287}.
\end{barticle}
\endbibitem

\bibitem[\protect\citeauthoryear{Higgins, Thompson and
  Spiegelhalter}{2009}]{higgins2009}
\begin{barticle}[author]
\bauthor{\bsnm{Higgins},~\bfnm{J.}\binits{J.}},
  \bauthor{\bsnm{Thompson},~\bfnm{S.~G.}\binits{S.~G.}} \AND
  \bauthor{\bsnm{Spiegelhalter},~\bfnm{D.~J.}\binits{D.~J.}}
(\byear{2009}).
\btitle{{A re-evaluation of random-effects meta-analysis}}.
\bjournal{Journal of the Royal Statistical Society: Series A (Statistics in
  Society)}
\bvolume{172}
\bpages{137--159}.
\bdoi{10.1111/j.1467-985x.2008.00552.x}
\end{barticle}
\endbibitem

\bibitem[\protect\citeauthoryear{Johnson}{2013}]{johnson2013vig}
\begin{btechreport}[author]
\bauthor{\bsnm{Johnson},~\bfnm{Toby}\binits{T.}}
(\byear{2013}).
\btitle{{Efficient calculation for multi-SNP genetic risk scores}}
\btype{Technical Report},
\bpublisher{The Comprehensive R Archive Network}.
\bnote{Available at
  http://cran.r-project.org/web/packages/gtx/vignettes/ashg2012.pdf [last
  accessed 2014/11/19]}.
\end{btechreport}
\endbibitem

\bibitem[\protect\citeauthoryear{Lawlor et~al.}{2008}]{lawlor2007}
\begin{barticle}[author]
\bauthor{\bsnm{Lawlor},~\bfnm{DA}\binits{D.}},
  \bauthor{\bsnm{Harbord},~\bfnm{RM}\binits{R.}},
  \bauthor{\bsnm{Sterne},~\bfnm{JAC}\binits{J.}},
  \bauthor{\bsnm{Timpson},~\bfnm{N}\binits{N.}} \AND
  \bauthor{\bsnm{Davey~Smith},~\bfnm{G}\binits{G.}}
(\byear{2008}).
\btitle{{Mendelian randomization: using genes as instruments for making causal
  inferences in epidemiology}}.
\bjournal{Statistics in Medicine}
\bvolume{27}
\bpages{1133--1163}.
\bdoi{10.1002/sim.3034}
\end{barticle}
\endbibitem

\bibitem[\protect\citeauthoryear{Martens et~al.}{2006}]{martens2006}
\begin{barticle}[author]
\bauthor{\bsnm{Martens},~\bfnm{E.~P.}\binits{E.~P.}},
  \bauthor{\bsnm{Pestman},~\bfnm{W.~R.}\binits{W.~R.}}, \bauthor{\bparticle{de}
  \bsnm{Boer},~\bfnm{A.}\binits{A.}},
  \bauthor{\bsnm{Belitser},~\bfnm{S.~V.}\binits{S.~V.}} \AND
  \bauthor{\bsnm{Klungel},~\bfnm{O.~H.}\binits{O.~H.}}
(\byear{2006}).
\btitle{{Instrumental variables: application and limitations}}.
\bjournal{Epidemiology}
\bvolume{17}
\bpages{260--267}.
\bdoi{10.1097/01.ede.0000215160.88317.cb}
\end{barticle}
\endbibitem

\bibitem[\protect\citeauthoryear{Nelson et~al.}{2015}]{nelson2015}
\begin{barticle}[author]
\bauthor{\bsnm{Nelson},~\bfnm{Christopher~P.}\binits{C.~P.}},
  \bauthor{\bsnm{Hamby},~\bfnm{Stephen~E.}\binits{S.~E.}},
  \bauthor{\bsnm{Saleheen},~\bfnm{Danish}\binits{D.}},
  \bauthor{\bsnm{Hopewell},~\bfnm{Jemma~C.}\binits{J.~C.}},
  \bauthor{\bsnm{Zeng},~\bfnm{Lingyao}\binits{L.}},
  \bauthor{\bsnm{Assimes},~\bfnm{Themistocles~L.}\binits{T.~L.}},
  \bauthor{\bsnm{Kanoni},~\bfnm{Stavroula}\binits{S.}},
  \bauthor{\bsnm{Willenborg},~\bfnm{Christina}\binits{C.}},
  \bauthor{\bsnm{Burgess},~\bfnm{Stephen}\binits{S.}},
  \bauthor{\bsnm{Amouyel},~\bfnm{Phillipe}\binits{P.}},
  \bauthor{\bsnm{Anand},~\bfnm{Sonia}\binits{S.}},
  \bauthor{\bsnm{Blankenberg},~\bfnm{Stefan}\binits{S.}},
  \bauthor{\bsnm{Boehm},~\bfnm{Bernhard~O.}\binits{B.~O.}},
  \bauthor{\bsnm{Clarke},~\bfnm{Robert~J.}\binits{R.~J.}},
  \bauthor{\bsnm{Collins},~\bfnm{Rory}\binits{R.}},
  \bauthor{\bsnm{Dedoussis},~\bfnm{George}\binits{G.}},
  \bauthor{\bsnm{Farrall},~\bfnm{Martin}\binits{M.}},
  \bauthor{\bsnm{Franks},~\bfnm{Paul~W.}\binits{P.~W.}},
  \bauthor{\bsnm{Groop},~\bfnm{Leif}\binits{L.}},
  \bauthor{\bsnm{Hall},~\bfnm{Alistair~S.}\binits{A.~S.}},
  \bauthor{\bsnm{Hamsten},~\bfnm{Anders}\binits{A.}},
  \bauthor{\bsnm{Hengstenberg},~\bfnm{Christian}\binits{C.}},
  \bauthor{\bsnm{Hovingh},~\bfnm{G.~Kees}\binits{G.~K.}},
  \bauthor{\bsnm{Ingelsson},~\bfnm{Erik}\binits{E.}},
  \bauthor{\bsnm{Kathiresan},~\bfnm{Sekar}\binits{S.}},
  \bauthor{\bsnm{Kee},~\bfnm{Frank}\binits{F.}},
  \bauthor{\bsnm{K\"{o}nig},~\bfnm{Inke~R.}\binits{I.~R.}},
  \bauthor{\bsnm{Kooner},~\bfnm{Jaspal}\binits{J.}},
  \bauthor{\bsnm{Lehtimäki},~\bfnm{Terho}\binits{T.}},
  \bauthor{\bsnm{M\"{a}rz},~\bfnm{Winifred}\binits{W.}},
  \bauthor{\bsnm{McPherson},~\bfnm{Ruth}\binits{R.}},
  \bauthor{\bsnm{Metspalu},~\bfnm{Andres}\binits{A.}},
  \bauthor{\bsnm{Nieminen},~\bfnm{Markku~S.}\binits{M.~S.}},
  \bauthor{\bsnm{O'Donnell},~\bfnm{Christopher~J.}\binits{C.~J.}},
  \bauthor{\bsnm{Palmer},~\bfnm{Colin N.~A.}\binits{C.~N.~A.}},
  \bauthor{\bsnm{Peters},~\bfnm{Annette}\binits{A.}},
  \bauthor{\bsnm{Perola},~\bfnm{Markus}\binits{M.}},
  \bauthor{\bsnm{Reilly},~\bfnm{Muredach~P.}\binits{M.~P.}},
  \bauthor{\bsnm{Ripatti},~\bfnm{Samuli}\binits{S.}},
  \bauthor{\bsnm{Roberts},~\bfnm{Robert}\binits{R.}},
  \bauthor{\bsnm{Salomaa},~\bfnm{Veikko}\binits{V.}},
  \bauthor{\bsnm{Shah},~\bfnm{Svati~H.}\binits{S.~H.}},
  \bauthor{\bsnm{Schreiber},~\bfnm{Stefan}\binits{S.}},
  \bauthor{\bsnm{Siegbahn},~\bfnm{Agneta}\binits{A.}},
  \bauthor{\bsnm{Thorsteinsdottir},~\bfnm{Unnur}\binits{U.}},
  \bauthor{\bsnm{Veronesi},~\bfnm{Giovani}\binits{G.}},
  \bauthor{\bsnm{Wareham},~\bfnm{Nicholas}\binits{N.}},
  \bauthor{\bsnm{Willer},~\bfnm{Cristen~J.}\binits{C.~J.}},
  \bauthor{\bsnm{Zalloua},~\bfnm{Pierre~A.}\binits{P.~A.}},
  \bauthor{\bsnm{Erdmann},~\bfnm{Jeanette}\binits{J.}},
  \bauthor{\bsnm{Deloukas},~\bfnm{Panos}\binits{P.}},
  \bauthor{\bsnm{Watkins},~\bfnm{Hugh}\binits{H.}},
  \bauthor{\bsnm{Schunkert},~\bfnm{Heribert}\binits{H.}},
  \bauthor{\bsnm{Danesh},~\bfnm{John}\binits{J.}},
  \bauthor{\bsnm{Thompson},~\bfnm{John~R.}\binits{J.~R.}} \AND
  \bauthor{\bsnm{Samani},~\bfnm{Nilesh~J.}\binits{N.~J.}}
(\byear{2015}).
\btitle{{Genetically determined height and coronary artery disease}}.
\bjournal{New England Journal of Medicine}
\bvolume{372}
\bpages{1608--1618}.
\bdoi{10.1056/NEJMoa1404881}
\end{barticle}
\endbibitem

\bibitem[\protect\citeauthoryear{Nitsch et~al.}{2006}]{nitsch2006}
\begin{barticle}[author]
\bauthor{\bsnm{Nitsch},~\bfnm{D.}\binits{D.}},
  \bauthor{\bsnm{Molokhia},~\bfnm{M.}\binits{M.}},
  \bauthor{\bsnm{Smeeth},~\bfnm{L.}\binits{L.}},
  \bauthor{\bsnm{DeStavola},~\bfnm{B.~L.}\binits{B.~L.}},
  \bauthor{\bsnm{Whittaker},~\bfnm{J.~C.}\binits{J.~C.}} \AND
  \bauthor{\bsnm{Leon},~\bfnm{D.~A.}\binits{D.~A.}}
(\byear{2006}).
\btitle{{Limits to causal inference based on Mendelian randomization: a
  comparison with randomized controlled trials}}.
\bjournal{American Journal of Epidemiology}
\bvolume{163}
\bpages{397--403}.
\bdoi{10.1093/aje/kwj062}
\end{barticle}
\endbibitem

\bibitem[\protect\citeauthoryear{Pearl}{2000}]{pearl2000}
\begin{bbook}[author]
\bauthor{\bsnm{Pearl},~\bfnm{J.}\binits{J.}}
(\byear{2000}).
\btitle{{Causality: models, reasoning, and inference}}.
\bpublisher{Cambridge University Press}.
\end{bbook}
\endbibitem

\bibitem[\protect\citeauthoryear{Pierce and Burgess}{2013}]{pierce2013}
\begin{barticle}[author]
\bauthor{\bsnm{Pierce},~\bfnm{B.}\binits{B.}} \AND
  \bauthor{\bsnm{Burgess},~\bfnm{S}\binits{S.}}
(\byear{2013}).
\btitle{{Efficient design for Mendelian randomization studies: subsample and
  two-sample instrumental variable estimators}}.
\bjournal{American Journal of Epidemiology}
\bvolume{178}
\bpages{1177--1184}.
\bdoi{10.1093/aje/kwt084}
\end{barticle}
\endbibitem

\bibitem[\protect\citeauthoryear{Riley et~al.}{2011}]{riley2011}
\begin{barticle}[author]
\bauthor{\bsnm{Riley},~\bfnm{Richard~D}\binits{R.~D.}},
  \bauthor{\bsnm{Higgins},~\bfnm{Julian~PT}\binits{J.~P.}},
  \bauthor{\bsnm{Deeks},~\bfnm{Jonathan~J}\binits{J.~J.}} \betal{et~al.}
(\byear{2011}).
\btitle{{Interpretation of random effects meta-analyses}}.
\bjournal{British Medical Journal}
\bvolume{342}
\bpages{d549}.
\bdoi{10.1136/bmj.d549}
\end{barticle}
\endbibitem

\bibitem[\protect\citeauthoryear{Schatzkin et~al.}{2009}]{schatzkin2009}
\begin{barticle}[author]
\bauthor{\bsnm{Schatzkin},~\bfnm{A.}\binits{A.}},
  \bauthor{\bsnm{Abnet},~\bfnm{C.~C.}\binits{C.~C.}},
  \bauthor{\bsnm{Cross},~\bfnm{A.~J.}\binits{A.~J.}},
  \bauthor{\bsnm{Gunter},~\bfnm{M.}\binits{M.}},
  \bauthor{\bsnm{Pfeiffer},~\bfnm{R.}\binits{R.}},
  \bauthor{\bsnm{Gail},~\bfnm{M.}\binits{M.}},
  \bauthor{\bsnm{Lim},~\bfnm{U.}\binits{U.}} \AND
  \bauthor{\bsnm{Davey~Smith},~\bfnm{G.}\binits{G.}}
(\byear{2009}).
\btitle{{Mendelian randomization: how it can -- and cannot -- help confirm
  causal relations between nutrition and cancer}}.
\bjournal{Cancer Prevention Research}
\bvolume{2}
\bpages{104--113}.
\bdoi{10.1158/1940-6207.capr-08-0070}
\end{barticle}
\endbibitem

\bibitem[\protect\citeauthoryear{Shen and Zhan}{2015}]{shen2015}
\begin{barticle}[author]
\bauthor{\bsnm{Shen},~\bfnm{Xia}\binits{X.}} \AND
  \bauthor{\bsnm{Zhan},~\bfnm{Yiqiang}\binits{Y.}}
(\byear{2015}).
\btitle{{Re: The effect on melanoma risk of genes previously associated with
  telomere length}}.
\bjournal{Journal of the National Cancer Institute}
\bvolume{107}
\bpages{djv237}.
\bdoi{10.1093/jnci/djv237}
\end{barticle}
\endbibitem

\bibitem[\protect\citeauthoryear{Staiger and Stock}{1997}]{staiger1997}
\begin{barticle}[author]
\bauthor{\bsnm{Staiger},~\bfnm{D}\binits{D.}} \AND
  \bauthor{\bsnm{Stock},~\bfnm{JH}\binits{J.}}
(\byear{1997}).
\btitle{{Instrumental variables regression with weak instruments}}.
\bjournal{Econometrica}
\bvolume{65}
\bpages{557--586}.
\end{barticle}
\endbibitem

\bibitem[\protect\citeauthoryear{Stock, Wright and Yogo}{2002}]{stock2002}
\begin{barticle}[author]
\bauthor{\bsnm{Stock},~\bfnm{JH}\binits{J.}},
  \bauthor{\bsnm{Wright},~\bfnm{JH}\binits{J.}} \AND
  \bauthor{\bsnm{Yogo},~\bfnm{M}\binits{M.}}
(\byear{2002}).
\btitle{{A survey of weak instruments and weak identification in generalized
  method of moments}}.
\bjournal{Journal of Business and Economic Statistics}
\bvolume{20}
\bpages{518--529}.
\bdoi{10.1198/073500102288618658}
\end{barticle}
\endbibitem

\bibitem[\protect\citeauthoryear{Stock and Yogo}{2002}]{stock2005}
\begin{barticle}[author]
\bauthor{\bsnm{Stock},~\bfnm{JH}\binits{J.}} \AND
  \bauthor{\bsnm{Yogo},~\bfnm{M}\binits{M.}}
(\byear{2002}).
\btitle{{Testing for weak instruments in linear IV regression}}.
\bjournal{SSRN eLibrary}
\bvolume{11}
\bpages{T0284}.
\end{barticle}
\endbibitem

\bibitem[\protect\citeauthoryear{Swanson and Hern{\'a}n}{2013}]{swanson2013}
\begin{barticle}[author]
\bauthor{\bsnm{Swanson},~\bfnm{Sonja}\binits{S.}} \AND
  \bauthor{\bsnm{Hern{\'a}n},~\bfnm{Miguel}\binits{M.}}
(\byear{2013}).
\btitle{{Commentary: how to report instrumental variable analyses (suggestions
  welcome)}}.
\bjournal{Epidemiology}
\bvolume{24}
\bpages{370--374}.
\bdoi{10.1097/ede.0b013e31828d0590}
\end{barticle}
\endbibitem

\bibitem[\protect\citeauthoryear{{R Core Team}}{2014}]{r312}
\begin{bmanual}[author]
\bauthor{\bsnm{{R Core Team}}}
(\byear{2014}).
\btitle{{R: A Language and Environment for Statistical Computing. Version 3.1.2
  (Pumpkin Helmet)}}
\bpublisher{R Foundation for Statistical Computing},
\baddress{Vienna, Austria}.
\end{bmanual}
\endbibitem

\bibitem[\protect\citeauthoryear{Thomas, Lawlor and
  Thompson}{2007}]{thomas2007}
\begin{barticle}[author]
\bauthor{\bsnm{Thomas},~\bfnm{D.~C.}\binits{D.~C.}},
  \bauthor{\bsnm{Lawlor},~\bfnm{D.~A.}\binits{D.~A.}} \AND
  \bauthor{\bsnm{Thompson},~\bfnm{J.~R.}\binits{J.~R.}}
(\byear{2007}).
\btitle{{Re: Estimation of bias in nongenetic observational studies using
  ``Mendelian triangulation'' by Bautista et al.}}
\bjournal{Annals of Epidemiology}
\bvolume{17}
\bpages{511--513}.
\bdoi{10.1016/j.annepidem.2006.12.005}
\end{barticle}
\endbibitem

\bibitem[\protect\citeauthoryear{Thompson and Sharp}{1999}]{thompson1999}
\begin{barticle}[author]
\bauthor{\bsnm{Thompson},~\bfnm{S.~G.}\binits{S.~G.}} \AND
  \bauthor{\bsnm{Sharp},~\bfnm{S.~J.}\binits{S.~J.}}
(\byear{1999}).
\btitle{{Explaining heterogeneity in meta-analysis: a comparison of methods}}.
\bjournal{Statistics in Medicine}
\bvolume{18}
\bpages{2693--2708}.
\end{barticle}
\endbibitem

\bibitem[\protect\citeauthoryear{VanderWeele and
  Hern{\'a}n}{2013}]{vanderweele2013}
\begin{barticle}[author]
\bauthor{\bsnm{VanderWeele},~\bfnm{TJ}\binits{T.}} \AND
  \bauthor{\bsnm{Hern{\'a}n},~\bfnm{MA}\binits{M.}}
(\byear{2013}).
\btitle{{Causal inference under multiple versions of treatment}}.
\bjournal{Journal of Causal Inference}
\bvolume{1}
\bpages{1--20}.
\bdoi{10.1515/jci-2012-0002}
\end{barticle}
\endbibitem

\bibitem[\protect\citeauthoryear{VanderWeele et~al.}{2014}]{vanderweele2014}
\begin{barticle}[author]
\bauthor{\bsnm{VanderWeele},~\bfnm{Tyler}\binits{T.}},
  \bauthor{\bsnm{Tchetgen~Tchetgen},~\bfnm{Eric}\binits{E.}},
  \bauthor{\bsnm{Cornelis},~\bfnm{Marilyn}\binits{M.}} \AND
  \bauthor{\bsnm{Kraft},~\bfnm{Peter}\binits{P.}}
(\byear{2014}).
\btitle{{Methodological challenges in Mendelian randomization}}.
\bjournal{Epidemiology}
\bvolume{25}
\bpages{427--435}.
\bdoi{10.1097/ede.0000000000000081}
\end{barticle}
\endbibitem

\bibitem[\protect\citeauthoryear{Zhang et~al.}{2015}]{zhang2015}
\begin{barticle}[author]
\bauthor{\bsnm{Zhang},~\bfnm{Chenan}\binits{C.}},
  \bauthor{\bsnm{Doherty},~\bfnm{Jennifer~A}\binits{J.~A.}},
  \bauthor{\bsnm{Burgess},~\bfnm{Stephen}\binits{S.}},
  \bauthor{\bsnm{Hung},~\bfnm{Rayjean~J}\binits{R.~J.}},
  \bauthor{\bsnm{Lindstr{\"o}m},~\bfnm{Sara}\binits{S.}},
  \bauthor{\bsnm{Kraft},~\bfnm{Peter}\binits{P.}},
  \bauthor{\bsnm{Gong},~\bfnm{Jian}\binits{J.}},
  \bauthor{\bsnm{Amos},~\bfnm{Christopher~I}\binits{C.~I.}},
  \bauthor{\bsnm{Sellers},~\bfnm{Thomas~A}\binits{T.~A.}},
  \bauthor{\bsnm{Monteiro},~\bfnm{Alvaro~NA}\binits{A.~N.}} \betal{et~al.}
(\byear{2015}).
\btitle{{Genetic determinants of telomere length and risk of common cancers: a
  Mendelian randomization study}}.
\bjournal{Human Molecular Genetics}.
\bdoi{10.1093/hmg/ddv252}
\end{barticle}
\endbibitem

\end{thebibliography}

\clearpage

\renewcommand{\thesection}{A.\arabic{section}}
\renewcommand{\thesubsection}{A.\arabic{subsection}}
\setcounter{table}{0}
\setcounter{figure}{0}
\setcounter{section}{0}
\setcounter{subsection}{0}
\setcounter{equation}{0}
\renewcommand{\theequation}{A\arabic{equation}}
\renewcommand{\thetable}{A\arabic{table}}
\renewcommand{\thefigure}{A\arabic{figure}}
\renewcommand{\tablename}{Appendix Table}
\renewcommand{\figurename}{Appendix Figure}
\section*{Appendix}
\subsection{Data for motivating example: causal effect of early menopause on triglycerides}
Information on the genetic variants included in the motivating analysis are presented in Appendix Table~\ref{summdata1}: for each variant, we provide the rsid, nearest gene(s), effect allele, other allele, association with early menopause (expressed as number of years earlier menopause) and standard error, and association with triglycerides (in standard deviation units) and standard error. Associations are also displayed visually as a scatter plot in Appendix Figure~\ref{weakscatter}. Associations with early menopause are obtained from Day et al.\ \citep{day2015}; larger numbers indicate that individuals with copies of the effect allele have earlier menopause on average compared with carriers of the other allele. These association estimates are available from download as part of the Supplementary Material to Day et al.\ (Supplementary Table 3). Associations with triglycerides are obtained from the Global Lipids Genetics Consortium \citep{willer2013}, and can be downloaded from \url{http://csg.sph.umich.edu//abecasis/public/lipids2013/}.

\setlength{\tabcolsep}{4pt}
\begin{table}[p]
\begin{footnotesize}
\begin{center}
\centering
\begin{tabular}[c]{ccccccccc}
\hline
Genetic     & Gene                       & Effect  & Other   & Early menopause     & Triglycerides       \\
variant     & region                     & allele  & allele  & in years (SE)       & SD difference (SE)  \\
\hline
rs10734411  & \emph{EIF3M}               & G       & A       & 0.12 (0.02)         &  0.0017 (0.0047)    \\
rs10852344  & \emph{GSPT1/BCAR4}         & T       & C       & 0.16 (0.02)         & -0.0030 (0.0047)    \\
rs10905065  & \emph{FBXO18}              & A       & G       & 0.11 (0.02)         & -0.0056 (0.0047)    \\
rs10957156  & \emph{CHD7}                & G       & A       & 0.14 (0.02)         &  0.0114 (0.0056)    \\
rs11031006  & \emph{FSHB}                & G       & A       & 0.25 (0.03)         & -0.0186 (0.0068)    \\
rs11668344  & \emph{BRSK1/NLRP11/U2AF2}  & A       & G       & 0.41 (0.02)         &  0.0009 (0.0049)    \\
rs11738223  & \emph{SH3PXD2B}            & G       & A       & 0.12 (0.02)         &  0.0007 (0.0036)    \\
rs1183272   & \emph{HELB}                & T       & C       & 0.31 (0.03)         &  0.0005 (0.0047)    \\
rs12142240  & \emph{RAD54L}              & C       & T       & 0.13 (0.02)         &  0.0051 (0.0050)    \\
rs12196873  & \emph{REV3L}               & A       & C       & 0.16 (0.03)         & -0.0099 (0.0068)    \\
rs12461110  & \emph{BRSK1/NLRP11/U2AF2}  & G       & A       & 0.15 (0.02)         &  0.0061 (0.0051)    \\
rs12824058  & \emph{PIWIL1}              & A       & G       & 0.14 (0.02)         &  0.0006 (0.0048)    \\
rs13040088  & \emph{SLCO4A1/DIDO1}       & A       & G       & 0.16 (0.02)         &  0.0004 (0.0057)    \\
rs1411478   & \emph{STX6}                & A       & G       & 0.13 (0.02)         & -0.0004 (0.0047)    \\
rs16858210  & \emph{PARL/POLR2H}         & A       & G       & 0.14 (0.02)         &  0.0023 (0.0055)    \\
rs16991615  & \emph{MCM8}                & A       & G       & 0.88 (0.04)         &  0.0025 (0.0073)    \\
rs1713460   & \emph{APEX1/PARP2/PNP}     & A       & G       & 0.14 (0.02)         &  0.0015 (0.0056)    \\
rs1799949   & \emph{BRCA1}               & A       & G       & 0.14 (0.02)         &  0.0107 (0.0049)    \\
rs1800932   & \emph{MSH6}                & G       & A       & 0.17 (0.03)         &  0.0020 (0.0060)    \\
rs2230365   & \emph{MSH5/HLA}            & T       & C       & 0.16 (0.03)         &  0.0202 (0.0046)    \\
rs2236553   & \emph{SLCO4A1/DIDO1}       & C       & T       & 0.16 (0.03)         & -0.0021 (0.0065)    \\
rs2241584   & \emph{UIMC1}               & A       & G       & 0.14 (0.02)         & -0.0007 (0.0048)    \\
rs2277339   & \emph{PRIM1/TAC3}          & G       & T       & 0.31 (0.03)         & -0.0072 (0.0080)    \\
rs2720044   & \emph{STAR}                & C       & A       & 0.29 (0.03)         &  0.0043 (0.0078)    \\
rs2941505   & \emph{STARD3/PGAP3/CDK12}  & A       & G       & 0.13 (0.02)         & -0.0074 (0.0035)    \\
rs349306    & \emph{POLR2E/KISS1R}       & G       & A       & 0.23 (0.04)         & -0.0082 (0.0055)    \\
rs365132    & \emph{UIMC1}               & G       & T       & 0.24 (0.02)         & -0.0003 (0.0047)    \\
rs3741604   & \emph{HELB}                & T       & C       & 0.29 (0.03)         & -0.0014 (0.0047)    \\
rs4246511   & \emph{RHBDL2/MYCBP}        & T       & C       & 0.22 (0.02)         &  0.0093 (0.0056)    \\
rs427394    & \emph{PAPD7}               & G       & A       & 0.13 (0.02)         & -0.0013 (0.0048)    \\
rs451417    & \emph{MCM8}                & C       & A       & 0.20 (0.03)         &  0.0019 (0.0081)    \\
rs4693089   & \emph{HELQ/FAM175A}        & G       & A       & 0.20 (0.02)         &  0.0045 (0.0048)    \\
rs4879656   & \emph{APTX}                & C       & A       & 0.12 (0.02)         &  0.0033 (0.0049)    \\
rs4886238   & \emph{TDRD3}               & A       & G       & 0.18 (0.02)         &  0.0009 (0.0050)    \\
rs551087    & \emph{SPPL3/SRSF9}         & A       & G       & 0.13 (0.02)         &  0.0032 (0.0036)    \\
rs5762534   & \emph{CHEK2}               & C       & T       & 0.16 (0.03)         &  0.0056 (0.0066)    \\
rs6484478   & \emph{FSHB}                & G       & A       & 0.14 (0.02)         & -0.0102 (0.0053)    \\
rs6856693   & \emph{ASCL1/MLF1IP}        & A       & G       & 0.16 (0.02)         & -0.0044 (0.0048)    \\
rs6899676   & \emph{SYCP2L/MAK}          & G       & A       & 0.21 (0.03)         &  0.0045 (0.0058)    \\
rs704795    & \emph{BRE/GTF3C2/EIFB4}    & G       & A       & 0.16 (0.02)         &  0.0567 (0.0034)    \\
rs707938    & \emph{MSH5/HLA}            & A       & G       & 0.16 (0.02)         &  0.0014 (0.0049)    \\
rs7259376   & \emph{ZNF729}              & A       & G       & 0.11 (0.02)         & -0.0041 (0.0047)    \\
rs763121    & \emph{DMC1/DDX17}          & G       & A       & 0.16 (0.02)         & -0.0179 (0.0036)    \\
rs8070740   & \emph{RPAIN}               & G       & A       & 0.15 (0.02)         &  0.0121 (0.0056)    \\
rs9039      & \emph{C16orf72/ABAT}       & C       & T       & 0.12 (0.02)         & -0.0068 (0.0037)    \\
rs930036    & \emph{TLK1/GAD1}           & A       & G       & 0.19 (0.02)         & -0.0001 (0.0049)    \\
rs9393800   & \emph{SYCP2L/MAK}          & A       & G       & 0.14 (0.02)         &  0.0073 (0.0054)    \\
\hline
\end{tabular}
\caption{List of genetic variants and associations (standard errors, SE) with early menopause (years) and with triglycerides (standard deviation, SD, difference) used in motivating example.} \label{summdata1}
\end{center}
\end{footnotesize}
\end{table}
\setlength{\tabcolsep}{6pt}

\begin{figure}[htbp]
\begin{center}
\includegraphics[width=10cm]{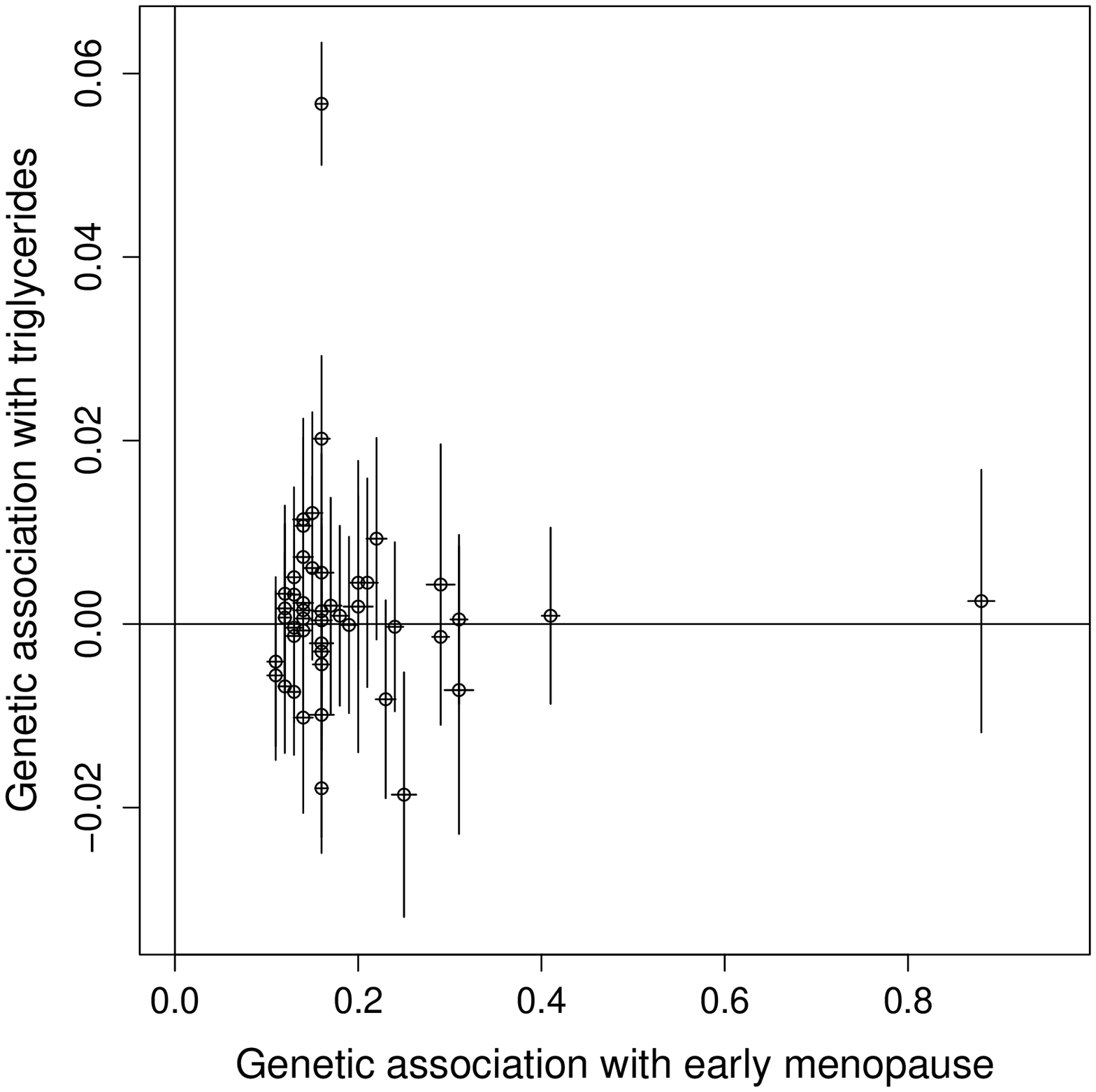}
\end{center}
\caption{Scatter plot of genetic associations with triglycerides (standard deviation units) against genetic associations with early menopause (in years).} \label{weakscatter}.
\end{figure}

\clearpage

\subsection{Code for implementing methods used in simulation study}
Code for performing the methods used in the simulation study for the R software package is provided below:

\scriptsize{
\begin{verbatim}
alpx=NULL; alpxsd=NULL                    # genetic associations with risk factor and standard errors
alpy=NULL; alpysd=NULL                    # genetic associations with outcome and standard errors

for (j in 1:vars) {
 alpx[j] = lm(x~g[,j])$coef[2]
 alpy[j] = lm(y~g[,j])$coef[2]
 alpxsd[j] = summary(lm(x~g[,j]))$coef[2,2]
 alpysd[j] = summary(lm(y~g[,j]))$coef[2,2]
}

 reg.first = summary(lm(alpy~alpx-1, weights=alpysd^-2))

 betafirst.fixed  = reg.first$coef[1]   # estimate using first-order weights, fixed-effect model
 betafirst.mulran = reg.first$coef[1]   # estimate using first-order weights, multiplicative random-effects
   sefirst.fixed  = reg.first$coef[1,2]/reg.first$sigma
                                        # standard error using first-order weights, fixed-effect model
   sefirst.mulran = reg.first$coef[1,2]/min(reg.first$sigma,1)
 betafirst.addran = metagen(alpy/alpx, abs(alpysd/alpx))$TE.random
                                        # estimate using first-order weights, additive random-effects model
   sefirst.addran = metagen(alpy/alpx, abs(alpysd/alpx))$seTE.random

 reg.second = summary(lm(alpy~alpx-1, weights=(alpysd^2+alpy^2*alpxsd^2/alpx^2)^-1))

 betasecond.fixed  = reg.second$coef[1] # estimate using second-order weights, fixed-effect model
 betasecond.mulran = reg.second$coef[1]
   sesecond.fixed  = reg.second$coef[1,2]/reg.second$sigma
   sesecond.mulran = reg.second$coef[1,2]/min(reg.second$sigma,1)
 betasecond.addran = metagen(alpy/alpx, sqrt(alpysd^2/alpx^2+alpy^2*alpxsd^2/alpx^4))$TE.random
   sesecond.addran = metagen(alpy/alpx, sqrt(alpysd^2/alpx^2+alpy^2*alpxsd^2/alpx^4))$seTE.random

theta = 0.1                             # correlation term from equation (1)

  reg.second.theta = summary(lm(alpy~alpx-1,
            weights=(alpysd^2+alpy^2*alpxsd^2/alpx^2-2*theta*alpy*alpxsd*alpysd/alpx)^-1))

 betasecond.theta.fixed  = reg.second.theta$coef[1]
                              # estimate using second-order weights with correlation, fixed-effect model
 betasecond.theta.mulran = reg.second.theta$coef[1]
   sesecond.theta.fixed  = reg.second.theta$coef[1,2]/reg.second.theta$sigma
   sesecond.theta.mulran = reg.second.theta$coef[1,2]/min(reg.second.theta$sigma,1)
 betasecond.theta.addran = metagen(alpy/alpx,
            sqrt(alpysd^2/alpx^2+alpy^2*alpxsd^2/alpx^4-2*theta*alpy*alpxsd*alpysd/alpx^3))$TE.random
   sesecond.theta.addran = metagen(alpy/alpx,
            sqrt(alpysd^2/alpx^2+alpy^2*alpxsd^2/alpx^4-2*theta*alpy*alpxsd*alpysd/alpx^3))$seTE.random
\end{verbatim}
}

\normalsize{
\hphantom{Send three- and four-pence, we're going to a dance.}
\clearpage

\subsection{Sensitivity analysis for value of $\theta$ in motivating example}
As stated in Section~\ref{sec:discuss:overlap}, in this paper we have assumed that the correlation parameter $\theta$ in the second-order expression for the variance of a causal estimate from the delta method (equation~\ref{secondvarwith}) is zero. While computational and practical considerations (the length of the simulation study to run, and the difficulty in estimating the parameter using summarized data only) preclude an investigation into the impact of this term in the simulation study, we can conduct a sensitivity analysis to consider the impact of the value of $\theta$ on estimates from the motivating example.

We conduct inverse-variance weighted analyses using weights derived from equation~(\ref{secondvarwith}) and fixed-effect, additive random-effects, and multiplicative random-effects models for $\theta = -0.2, -0.1, 0, 0.1, 0.2, 0.3$. The causal estimates and 95\% confidence intervals from each analysis are presented in Appendix Table~\ref{results9}. We see that the estimates and confidence intervals do not change substantially despite the wide range of values of $\theta$ considered.

The true value of $\theta$ should be zero if the associations with the risk factor and outcome are estimated in non-overlapping samples, and similar to the correlation between the risk factor and the outcome if the associations are estimated in the same individuals. With partial overlap, the value of $\theta$ will be between these two values.

\setlength{\tabcolsep}{5pt}
\begin{table}[h]
\begin{footnotesize}
\begin{center}
\centering
\begin{tabular}[c]{c|cc|cc|cc}
\hline
         & \multicolumn{2}{c|}{Fixed-effects} & \multicolumn{2}{c|}{Additive random-effects} & \multicolumn{2}{c}{Multiplicative random-effects} \\
\cline{2-7}
$\theta$ & Estimate  & 95\% CI                & Estimate  & 95\% CI                          &  Estimate  & 95\% CI                              \\
\hline
$-0.2$   &  0.000    & -0.007, 0.007          &  0.004    & -0.010, 0.017                    &  0.000    & -0.012, 0.012                         \\
$-0.1$   &  0.001    & -0.006, 0.008          &  0.005    & -0.009, 0.018                    &  0.001    & -0.011, 0.013                         \\
0        &  0.002    & -0.005, 0.009          &  0.006    & -0.008, 0.019                    &  0.002    & -0.010, 0.014                         \\
0.1      &  0.003    & -0.004, 0.011          &  0.007    & -0.007, 0.020                    &  0.003    & -0.009, 0.016                         \\
0.2      &  0.004    & -0.003, 0.012          &  0.008    & -0.006, 0.022                    &  0.004    & -0.008, 0.017                         \\
0.3      &  0.005    & -0.002, 0.013          &  0.009    & -0.006, 0.023                    &  0.005    & -0.008, 0.018                         \\
\hline
\end{tabular}
\caption{Estimates and 95\% confidence intervals (CI) from inverse-variance weighted analyses using second-order weights from equation~(\ref{secondvarwith}) with different values of the correlation parameter $\theta$.} \label{results9}
\end{center}
\end{footnotesize}
\end{table}
\setlength{\tabcolsep}{6pt}

\clearpage

\subsection{Additional results from simulation study}
Additional results from scenarios 1 and 2 are presented in Appendix Table~\ref{results.1a}, and from scenarios 3 and 4 in Appendix Table~\ref{results.3a}. For each value of the instrument strength, the (Monte Carlo) standard deviation and the mean standard error of estimates are presented. Using second-order weights, only results from the fixed-effect analyses are presented, as heterogeneity was not detected in the vast majority of datasets, and so results were the same up to 3 decimal places in almost all cases.

\setlength{\tabcolsep}{5pt}
\begin{table}[hbtp]
\begin{minipage}{\textwidth}
\begin{adjustwidth}{-0.5cm}{-0.5cm}
\begin{footnotesize}
\begin{center}
\centering
\begin{tabular}[c]{c|cc|cc|cc|cc}
\hline
         & \multicolumn{2}{c|}{Second-order}            &  \multicolumn{6}{c}{First-order weights}                         \\
\cline{2-9}
         & \multicolumn{2}{c|}{Fixed-effect}            &  \multicolumn{2}{c|}{Fixed-effect}  &
           \multicolumn{2}{c|}{Additive random-effects} &  \multicolumn{2}{c}{Multiplicative random-effects}               \\
\hline
$\alpha$ &  SD   &  Mean SE   &  SD   &  Mean SE   &  \hspace{2mm} SD  \hspace{2mm}  &  Mean SE   & \hspace{3mm} SD \hspace{3mm}   & \hspace{3mm}  Mean SE \hspace{3mm} \\
\hline
\multicolumn{9}{c}{Scenario 1. One-sample setting, valid instrumental variables}           \\
\hline
\multicolumn{9}{c}{1a. Null causal effect ($\beta_X = 0$), positive confounding ($\beta_U = +1$).}                         \\
\hline
0.03     & 0.117 &  0.163     & 0.140 &  0.147     & 0.142 &  0.159     & 0.140 &  0.153   \\
0.05     & 0.093 &  0.121     & 0.107 &  0.113     & 0.110 &  0.122     & 0.107 &  0.118   \\
0.08     & 0.067 &  0.080     & 0.076 &  0.077     & 0.078 &  0.082     & 0.076 &  0.081   \\
0.10     & 0.056 &  0.065     & 0.062 &  0.063     & 0.064 &  0.067     & 0.062 &  0.066   \\
\hline
\multicolumn{9}{c}{1b. Null causal effect ($\beta_X = 0$), negative confounding ($\beta_U = -1$).}                         \\
\hline
0.03     & 0.119 &  0.165     & 0.141 &  0.149     & 0.144 &  0.160     & 0.141 &  0.155   \\
0.05     & 0.091 &  0.120     & 0.106 &  0.112     & 0.108 &  0.121     & 0.106 &  0.117   \\
0.08     & 0.067 &  0.080     & 0.075 &  0.077     & 0.077 &  0.083     & 0.075 &  0.081   \\
0.10     & 0.056 &  0.065     & 0.062 &  0.063     & 0.063 &  0.067     & 0.062 &  0.066   \\
\hline
\multicolumn{9}{c}{1c. Positive causal effect ($\beta_X = +0.2$), positive confounding ($\beta_U = +1$).}                  \\
\hline
0.03     & 0.122 &  0.185     & 0.140 &  0.165     & 0.141 &  0.169     & 0.140 &  0.167   \\
0.05     & 0.096 &  0.137     & 0.107 &  0.125     & 0.108 &  0.129     & 0.107 &  0.127   \\
0.08     & 0.069 &  0.093     & 0.075 &  0.088     & 0.076 &  0.090     & 0.075 &  0.089   \\
0.10     & 0.059 &  0.072     & 0.063 &  0.069     & 0.064 &  0.071     & 0.063 &  0.070   \\
\hline
\multicolumn{9}{c}{1d. Positive causal effect ($\beta_X = +0.2$), negative confounding ($\beta_U = -1$).}                  \\
\hline
0.03     & 0.115 &  0.150     & 0.140 &  0.136     & 0.145 &  0.157     & 0.140 &  0.146   \\
0.05     & 0.088 &  0.107     & 0.105 &  0.099     & 0.110 &  0.115     & 0.105 &  0.109   \\
0.08     & 0.066 &  0.075     & 0.075 &  0.071     & 0.078 &  0.082     & 0.075 &  0.079   \\
0.10     & 0.056 &  0.063     & 0.063 &  0.060     & 0.066 &  0.069     & 0.063 &  0.067   \\
\hline
\hline
\multicolumn{9}{c}{Scenario 2. One-sample setting, invalid instrumental variables}                                         \\
\hline
\multicolumn{9}{c}{2a. Null causal effect ($\beta_X = 0$), positive confounding ($\beta_U = +1$).}                         \\
\hline
0.03     & 0.136 &  0.167     & 0.170 &  0.147     & 0.183 &  0.192     & 0.170 &  0.171   \\
0.05     & 0.106 &  0.121     & 0.128 &  0.110     & 0.137 &  0.143     & 0.128 &  0.130   \\
0.08     & 0.079 &  0.084     & 0.092 &  0.079     & 0.098 &  0.100     & 0.092 &  0.094   \\
0.10     & 0.067 &  0.067     & 0.076 &  0.064     & 0.080 &  0.080     & 0.076 &  0.077   \\
\hline
\multicolumn{9}{c}{2b. Null causal effect ($\beta_X = 0$), negative confounding ($\beta_U = -1$).}                         \\
\hline
0.03     & 0.136 &  0.166     & 0.169 &  0.147     & 0.182 &  0.193     & 0.169 &  0.172   \\
0.05     & 0.108 &  0.120     & 0.130 &  0.110     & 0.139 &  0.142     & 0.130 &  0.130   \\
0.08     & 0.078 &  0.081     & 0.091 &  0.077     & 0.096 &  0.096     & 0.091 &  0.091   \\
0.10     & 0.067 &  0.068     & 0.076 &  0.065     & 0.080 &  0.080     & 0.076 &  0.077   \\
\hline
\multicolumn{9}{c}{2c. Positive causal effect ($\beta_X = +0.2$), positive confounding ($\beta_U = +1$).}                  \\
\hline
0.03     & 0.144 &  0.188     & 0.172 &  0.164     & 0.179 &  0.192     & 0.172 &  0.178   \\
0.05     & 0.111 &  0.136     & 0.129 &  0.123     & 0.134 &  0.143     & 0.129 &  0.135   \\
0.08     & 0.083 &  0.095     & 0.093 &  0.088     & 0.096 &  0.101     & 0.093 &  0.097   \\
0.10     & 0.069 &  0.077     & 0.076 &  0.072     & 0.078 &  0.082     & 0.076 &  0.080   \\
\hline
\multicolumn{9}{c}{2d. Positive causal effect ($\beta_X = +0.2$), negative confounding ($\beta_U = -1$).}                  \\
\hline
0.03     & 0.134 &  0.153     & 0.172 &  0.135     & 0.191 &  0.198     & 0.172 &  0.170   \\
0.05     & 0.106 &  0.114     & 0.131 &  0.103     & 0.144 &  0.148     & 0.131 &  0.132   \\
0.08     & 0.077 &  0.075     & 0.091 &  0.070     & 0.098 &  0.097     & 0.091 &  0.091   \\
0.10     & 0.065 &  0.060     & 0.074 &  0.057     & 0.079 &  0.077     & 0.074 &  0.073   \\
\hline
\end{tabular}
\caption{\small{Further simulation study results for scenarios 1 and 2 (one-sample setting, valid and invalid instrumental variables): standard deviation (SD) of estimates and mean standard error (SE) for various inverse-variance weighted methods with four sets of parameter values (null and positive causal effect, positive and negative confounding) for different strengths of instrument ($\alpha$).}} \label{results.1a}
\end{center}
\end{footnotesize} 
\end{adjustwidth}
\end{minipage}
\end{table}
\setlength{\tabcolsep}{6pt}

\setlength{\tabcolsep}{5pt}
\begin{table}[hbtp]
\begin{minipage}{\textwidth}
\begin{adjustwidth}{-0.5cm}{-0.5cm}
\begin{footnotesize}
\begin{center}
\centering
\begin{tabular}[c]{c|cc|cc|cc|cc}
\hline
         & \multicolumn{2}{c|}{Second-order}            &  \multicolumn{6}{c}{First-order weights}                         \\
\cline{2-9}
         & \multicolumn{2}{c|}{Fixed-effect}            &  \multicolumn{2}{c|}{Fixed-effect}  &
           \multicolumn{2}{c|}{Additive random-effects} &  \multicolumn{2}{c}{Multiplicative random-effects}               \\
\hline
$\alpha$ &  SD   &  Mean SE   &  SD   &  Mean SE   &  \hspace{2mm} SD  \hspace{2mm}  &  Mean SE   & \hspace{3mm} SD \hspace{3mm}   & \hspace{3mm}  Mean SE \hspace{3mm} \\
\hline
\multicolumn{9}{c}{Scenario 3. Two-sample setting, valid instrumental variables}           \\
\hline
\multicolumn{9}{c}{3a. Null causal effect ($\beta_X = 0$), positive confounding ($\beta_U = +1$).}                         \\
\hline
0.03     & 0.123 &  0.162     & 0.150 &  0.148     & 0.154 &  0.165     & 0.150 &  0.156   \\
0.05     & 0.093 &  0.117     & 0.108 &  0.109     & 0.110 &  0.120     & 0.108 &  0.116   \\
0.08     & 0.067 &  0.079     & 0.076 &  0.076     & 0.076 &  0.082     & 0.076 &  0.081   \\
0.10     & 0.057 &  0.066     & 0.063 &  0.064     & 0.063 &  0.069     & 0.063 &  0.067   \\
\hline
\multicolumn{9}{c}{3b. Null causal effect ($\beta_X = 0$), negative confounding ($\beta_U = -1$).}                         \\
\hline
0.03     & 0.124 &  0.163     & 0.150 &  0.148     & 0.154 &  0.167     & 0.150 &  0.157   \\
0.05     & 0.095 &  0.118     & 0.111 &  0.111     & 0.113 &  0.122     & 0.111 &  0.117   \\
0.08     & 0.069 &  0.082     & 0.078 &  0.078     & 0.079 &  0.085     & 0.078 &  0.083   \\
0.10     & 0.059 &  0.067     & 0.065 &  0.064     & 0.066 &  0.069     & 0.065 &  0.068   \\
\hline
\multicolumn{9}{c}{3c. Positive causal effect ($\beta_X = +0.2$), positive confounding ($\beta_U = +1$).}                  \\
\hline
0.03     & 0.138 &  0.182     & 0.168 &  0.165     & 0.173 &  0.185     & 0.168 &  0.175   \\
0.05     & 0.108 &  0.135     & 0.127 &  0.125     & 0.130 &  0.139     & 0.127 &  0.133   \\
0.08     & 0.080 &  0.093     & 0.090 &  0.088     & 0.091 &  0.096     & 0.090 &  0.094   \\
0.10     & 0.065 &  0.073     & 0.071 &  0.070     & 0.072 &  0.076     & 0.071 &  0.075   \\
\hline
\multicolumn{9}{c}{3d. Positive causal effect ($\beta_X = +0.2$), negative confounding ($\beta_U = -1$).}                  \\
\hline
0.03     & 0.113 &  0.149     & 0.140 &  0.135     & 0.144 &  0.153     & 0.140 &  0.144   \\
0.05     & 0.089 &  0.111     & 0.104 &  0.102     & 0.106 &  0.115     & 0.104 &  0.109   \\
0.08     & 0.065 &  0.076     & 0.073 &  0.072     & 0.074 &  0.079     & 0.073 &  0.077   \\
0.10     & 0.054 &  0.061     & 0.059 &  0.058     & 0.060 &  0.063     & 0.059 &  0.062   \\
\hline
\hline
\multicolumn{9}{c}{Scenario 4. Two-sample setting, invalid instrumental variables}                                         \\
\hline
\multicolumn{9}{c}{4a. Null causal effect ($\beta_X = 0$), positive confounding ($\beta_U = +1$).}                         \\
\hline
0.03     & 0.144 &  0.166     & 0.180 &  0.147     & 0.193 &  0.202     & 0.180 &  0.177   \\
0.05     & 0.108 &  0.121     & 0.131 &  0.111     & 0.138 &  0.147     & 0.131 &  0.133   \\
0.08     & 0.081 &  0.083     & 0.094 &  0.078     & 0.098 &  0.100     & 0.094 &  0.094   \\
0.10     & 0.065 &  0.064     & 0.074 &  0.062     & 0.076 &  0.077     & 0.074 &  0.074   \\
\hline
\multicolumn{9}{c}{4b. Null causal effect ($\beta_X = 0$), negative confounding ($\beta_U = -1$).}                         \\
\hline
0.03     & 0.143 &  0.169     & 0.181 &  0.150     & 0.196 &  0.204     & 0.181 &  0.180   \\
0.05     & 0.110 &  0.121     & 0.134 &  0.111     & 0.141 &  0.147     & 0.134 &  0.133   \\
0.08     & 0.080 &  0.083     & 0.094 &  0.078     & 0.097 &  0.099     & 0.094 &  0.094   \\
0.10     & 0.068 &  0.066     & 0.077 &  0.063     & 0.079 &  0.079     & 0.077 &  0.076   \\
\hline
\multicolumn{9}{c}{4c. Positive causal effect ($\beta_X = +0.2$), positive confounding ($\beta_U = +1$).}                  \\
\hline
0.03     & 0.156 &  0.184     & 0.194 &  0.164     & 0.209 &  0.217     & 0.194 &  0.193   \\
0.05     & 0.121 &  0.136     & 0.146 &  0.124     & 0.154 &  0.160     & 0.146 &  0.146   \\
0.08     & 0.087 &  0.092     & 0.101 &  0.086     & 0.105 &  0.107     & 0.101 &  0.102   \\
0.10     & 0.073 &  0.074     & 0.083 &  0.070     & 0.085 &  0.086     & 0.083 &  0.083   \\
\hline
\multicolumn{9}{c}{4d. Positive causal effect ($\beta_X = +0.2$), negative confounding ($\beta_U = -1$).}                  \\
\hline
0.03     & 0.135 &  0.154     & 0.172 &  0.135     & 0.188 &  0.193     & 0.172 &  0.168   \\
0.05     & 0.105 &  0.112     & 0.128 &  0.102     & 0.136 &  0.141     & 0.128 &  0.127   \\
0.08     & 0.074 &  0.074     & 0.087 &  0.069     & 0.091 &  0.092     & 0.087 &  0.086   \\
0.10     & 0.063 &  0.061     & 0.072 &  0.058     & 0.074 &  0.075     & 0.072 &  0.072   \\
\hline
\end{tabular}
\caption{\small{Further simulation study results for scenarios 3 and 4 (two-sample setting, valid and invalid instrumental variables): standard deviation (SD) of estimates and mean standard error (SE) for various inverse-variance weighted methods with four sets of parameter values (null and positive causal effect, positive and negative confounding) for different strengths of instrument ($\alpha$).}} \label{results.3a}
\end{center}
\end{footnotesize} 
\end{adjustwidth}
\end{minipage}
\end{table}
\setlength{\tabcolsep}{6pt}

\clearpage

\subsection{Additional simulation with directional pleiotropy}
To provide some guidance as to the performance of the inverse-variance weighted method when there is directional pleiotropy, we perform a further simulation under this scenario. The parameters and scenarios are taken to be the same as those in the main body of the paper, except that rather than drawing the genetic effects on the risk factor ($\alpha_j$) and the direct effects of the genetic variants on the outcome ($\beta_{Zj}$) from independent normal distributions as in Scenarios 2 and 4, we draw them from a bivariate normal distribution. The univariate distributions of these parameters are the same (the $\alpha_j$ parameters have mean $\alpha$ and variance $0.02^2$; the $\beta_{Zj}$ parameters have mean 0 and variance $0.02^2$), but the correlation between the distributions is set to 0.4.
\begin{equation}
\left(\begin{array}{c} \alpha_j \\ \beta_{Zj} \end{array} \right) \sim \mathcal{N}_2 \Big( \left(\begin{array}{c} \alpha \\ 0 \end{array} \right), \left(\begin{array}{cc} 0.02^2 & 0.4 \times 0.02^2 \\ 0.4 \times 0.02^2 & 0.02^2 \end{array} \right) \Big) \notag
\end{equation}
This correlation means that the direct effects of genetic variants on the outcome are greater for those variants that have stronger effects on the risk factor, and so for those variants that receive more weight in the analysis. Hence, although the overall mean pleiotropic effect has mean zero, pleiotropic effects of weak and strong instruments separately do not have mean zero. We refer to the one-sample setting with directional pleiotropy as Scenario 6, and the two-sample setting with directional pleiotropy as Scenario 7.

Results for the mean estimate and empirical power to detect a causal effect are given in Appendix Table~\ref{results.5}. In the one-sample setting (scenario 6), there is bias in the direction of confounding in all cases. While Type 1 error rates under the null are inflated throughout, there is a clear preference for the use of second-order weights and random-effects models, as well as a slight preference for the additive random-effects model (based on slightly more conservative coverage properties with first-order weights). This mirrors the advice in the main paper. In the two-sample setting (scenario 7), bias under the null is in the positive direction, whereas bias under the alternative is towards the null. Type 1 error rates under the null with random-effects models are close to nominal levels, with conservative coverage for second-order weights, and slightly anti-conservative coverage for first-order weights. However, the advice from the main paper to use first-order weights in a two-sample setting would not lead to overly misleading inferences, as Type 1 error rates with first-order weights are close to the nominal 5\% level. Power to detect a causal effect is greater using first-order weights in this case. Hence, on the basis of these simulations, the advice in the main body of the paper also holds with directional pleiotropy.

In practice, we repeat that estimates from the inverse-variance weighted method will typically be biased if the genetic variants are not valid instruments (and the example of directional pleiotropy considered here is far from extreme), and recommend the use of robust methods (such as the Egger method and median-based methods introduced in the discussion of the paper) as sensitivity analyses for applied Mendelian randomization investigations.

\setlength{\tabcolsep}{4pt} 
\begin{table}[hbtp]
\begin{minipage}{\textwidth}
\begin{adjustwidth}{-0.5cm}{-0.5cm}
\begin{footnotesize}
\begin{center}
\centering
\begin{tabular}[c]{ccc|cc|cc|cc|cc|cc|cc}
\hline
\multicolumn{3}{c|}{}         & \multicolumn{4}{c|}{Fixed-effect}  & \multicolumn{4}{c|}{Additive random-effects} & \multicolumn{4}{c}{Multiplicative random-effects} \\
\cline{4-15}
\multicolumn{3}{c|}{}         & \multicolumn{2}{c|}{Second-order}  &  \multicolumn{2}{c|}{First-order} &
                                \multicolumn{2}{c|}{Second-order}  &  \multicolumn{2}{c|}{First-order} &
                                \multicolumn{2}{c|}{Second-order}  &  \multicolumn{2}{c}{First-order}  \\
\hline
$\alpha$ &  $F$     &  $R^2$  &  Mean  &  Power  &  Mean  &  Power  &  Mean  &  Power  &  Mean  &  Power  &  Mean  &  Power  &  Mean  &  Power  \\
\hline
\multicolumn{15}{c}{Scenario 6. One-sample setting, invalid instrumental variables with directional pleiotropy}                                 \\
\hline
\multicolumn{15}{c}{6a. Null causal effect ($\beta_X = 0$), positive confounding ($\beta_U = +1$).}                                             \\
\hline
0.03     &  2.4     & 1.0\%   &  0.218 &  22.1   &  0.275 &  48.4   &  0.218 &  22.1   &  0.262 &  32.3   &  0.218 &  22.1   &  0.275 &  39.7   \\
0.05     &  4.2     & 1.7\%   &  0.142 &  20.1   &  0.152 &  32.3   &  0.142 &  20.0   &  0.123 &  17.4   &  0.142 &  20.1   &  0.152 &  23.9   \\
0.08     &  8.6     & 3.3\%   &  0.080 &  16.0   &  0.073 &  20.6   &  0.078 &  14.6   &  0.044 &  10.0   &  0.080 &  15.2   &  0.073 &  13.9   \\
0.10     & 12.5     & 4.8\%   &  0.056 &  14.8   &  0.048 &  18.1   &  0.051 &  11.7   &  0.025 &   8.9   &  0.056 &  12.9   &  0.048 &  11.7   \\
\hline
\multicolumn{15}{c}{6b. Null causal effect ($\beta_X = 0$), negative confounding ($\beta_U = -1$).}                                             \\
\hline
0.03     &  2.4     & 1.0\%   & -0.098 &   5.7   & -0.125 &  19.2   & -0.098 &   5.7   & -0.118 &  10.1   & -0.098 &   5.7   & -0.125 &  12.8   \\
0.05     &  4.2     & 1.7\%   & -0.069 &   6.5   & -0.076 &  15.6   & -0.069 &   6.4   & -0.060 &   7.2   & -0.069 &   6.4   & -0.076 &   9.5   \\
0.08     &  8.6     & 3.3\%   & -0.038 &   6.8   & -0.034 &  12.8   & -0.036 &   6.2   & -0.019 &   6.1   & -0.038 &   6.4   & -0.034 &   7.6   \\
0.10     & 12.5     & 4.8\%   & -0.027 &   6.3   & -0.024 &  10.5   & -0.024 &   5.2   & -0.012 &   5.4   & -0.027 &   5.5   & -0.024 &   6.3   \\
\hline
\multicolumn{15}{c}{6c. Positive causal effect ($\beta_X = +0.2$), positive confounding ($\beta_U = +1$).}                                      \\
\hline
0.03     &  2.4     & 1.0\%   &  0.380 &  55.6   &  0.471 &  81.6   &  0.380 &  55.6   &  0.465 &  73.3   &  0.380 &  55.6   &  0.471 &  78.4   \\
0.05     &  4.2     & 1.7\%   &  0.309 &  66.2   &  0.351 &  80.2   &  0.309 &  66.1   &  0.334 &  67.8   &  0.309 &  66.1   &  0.351 &  75.8   \\
0.08     &  8.6     & 3.3\%   &  0.255 &  81.3   &  0.275 &  86.4   &  0.255 &  80.7   &  0.257 &  74.5   &  0.255 &  81.1   &  0.275 &  82.5   \\
0.10     & 12.5     & 4.8\%   &  0.232 &  89.2   &  0.247 &  91.7   &  0.230 &  87.5   &  0.232 &  82.4   &  0.232 &  88.6   &  0.247 &  88.5   \\
\hline
\multicolumn{15}{c}{6d. Positive causal effect ($\beta_X = +0.2$), negative confounding ($\beta_U = -1$).}                                      \\
\hline
0.03     &  2.4     & 1.0\%   &  0.047 &   3.5   &  0.073 &  16.4   &  0.047 &   3.5   &  0.083 &   7.0   &  0.047 &   3.5   &  0.073 &   8.2   \\
0.05     &  4.2     & 1.7\%   &  0.086 &  10.4   &  0.126 &  28.8   &  0.086 &  10.3   &  0.146 &  16.1   &  0.086 &  10.3   &  0.126 &  17.0   \\
0.08     &  8.6     & 3.4\%   &  0.130 &  42.6   &  0.168 &  63.6   &  0.134 &  40.2   &  0.186 &  49.6   &  0.130 &  39.8   &  0.168 &  47.1   \\
0.10     & 12.5     & 4.8\%   &  0.144 &  64.3   &  0.177 &  79.9   &  0.151 &  60.0   &  0.193 &  69.0   &  0.144 &  58.1   &  0.177 &  64.6   \\
\hline
\hline
\multicolumn{15}{c}{Scenario 7. Two-sample setting, invalid instrumental variables with directional pleiotropy}                                 \\
\hline
\multicolumn{15}{c}{7a. Null causal effect ($\beta_X = 0$), positive confounding ($\beta_U = +1$).}                                             \\
\hline
0.03     &  2.4     & 1.0\%   &  0.056 &   3.1   &  0.074 &  13.1   &  0.056 &   3.1   &  0.071 &   6.1   &  0.056 &   3.1   &  0.074 &   8.0   \\
0.05     &  4.2     & 1.7\%   &  0.037 &   4.0   &  0.041 &  11.6   &  0.037 &   4.0   &  0.032 &   5.2   &  0.037 &   4.0   &  0.041 &   6.6   \\
0.08     &  8.6     & 3.3\%   &  0.022 &   5.3   &  0.020 &  11.8   &  0.021 &   4.8   &  0.012 &   5.8   &  0.022 &   4.9   &  0.020 &   6.7   \\
0.10     & 12.5     & 4.8\%   &  0.017 &   5.6   &  0.014 &  10.5   &  0.015 &   4.6   &  0.007 &   5.4   &  0.017 &   4.8   &  0.014 &   6.1   \\
\hline
\multicolumn{15}{c}{7b. Null causal effect ($\beta_X = 0$), negative confounding ($\beta_U = -1$).}                                             \\
\hline
0.03     &  2.4     & 1.0\%   &  0.056 &   2.7   &  0.072 &  13.0   &  0.056 &   2.7   &  0.068 &   6.0   &  0.056 &   2.7   &  0.072 &   7.5   \\
0.05     &  4.3     & 1.7\%   &  0.037 &   3.9   &  0.040 &  12.5   &  0.037 &   3.9   &  0.032 &   5.5   &  0.037 &   3.9   &  0.040 &   7.1   \\
0.08     &  8.6     & 3.3\%   &  0.022 &   5.1   &  0.020 &  11.1   &  0.021 &   4.7   &  0.012 &   5.6   &  0.022 &   4.8   &  0.020 &   6.5   \\
0.10     & 12.5     & 4.8\%   &  0.017 &   5.8   &  0.015 &  11.0   &  0.016 &   5.0   &  0.008 &   5.8   &  0.017 &   5.2   &  0.015 &   6.4   \\
\hline
\multicolumn{15}{c}{7c. Positive causal effect ($\beta_X = +0.2$), positive confounding ($\beta_U = +1$).}                                      \\
\hline
0.03     &  2.4     & 1.0\%   &  0.141 &   7.6   &  0.192 &  26.5   &  0.141 &   7.6   &  0.193 &  15.1   &  0.141 &   7.6   &  0.192 &  18.3   \\
0.05     &  4.2     & 1.7\%   &  0.159 &  19.7   &  0.203 &  40.5   &  0.159 &  19.6   &  0.204 &  26.5   &  0.159 &  19.6   &  0.203 &  29.9   \\
0.08     &  8.6     & 3.3\%   &  0.164 &  43.4   &  0.196 &  61.1   &  0.165 &  41.6   &  0.197 &  46.3   &  0.164 &  41.9   &  0.196 &  49.2   \\
0.10     & 12.5     & 4.8\%   &  0.171 &  63.9   &  0.197 &  76.7   &  0.173 &  60.1   &  0.199 &  63.3   &  0.171 &  60.5   &  0.197 &  65.7   \\
\hline
\multicolumn{15}{c}{7d. Positive causal effect ($\beta_X = +0.2$), negative confounding ($\beta_U = -1$).}                                      \\
\hline
0.03     &  2.4     & 1.0\%   &  0.137 &  11.1   &  0.189 &  34.4   &  0.137 &  11.1   &  0.188 &  17.7   &  0.137 &  11.1   &  0.189 &  22.0   \\
0.05     &  4.2     & 1.7\%   &  0.150 &  24.9   &  0.196 &  48.9   &  0.150 &  24.7   &  0.197 &  30.0   &  0.150 &  24.7   &  0.196 &  35.0   \\
0.08     &  8.6     & 3.3\%   &  0.161 &  56.9   &  0.195 &  73.8   &  0.163 &  53.5   &  0.196 &  56.0   &  0.161 &  54.0   &  0.195 &  59.6   \\
0.10     & 12.5     & 4.7\%   &  0.169 &  75.5   &  0.198 &  86.5   &  0.171 &  70.4   &  0.200 &  72.3   &  0.169 &  70.7   &  0.198 &  75.2   \\
\hline
\end{tabular}
\caption{\small{Simulation study results for scenarios 6 and 7 (invalid instrumental variables with directional pleiotropy, one-sample and two-sample settings): mean estimate and power (\%) of 95\% confidence interval for various inverse-variance weighted methods with four sets of parameter values (null and positive causal effect, positive and negative confounding. The strength of the genetic variants as instruments is indicated: mean per allele effect on the risk factor ($\alpha$), mean F statistic ($F$) and mean coefficient of determination ($R^2$) from regression of risk factor on the genetic variants.}} \label{results.5}
\end{center}
\end{footnotesize} 
\end{adjustwidth}
\end{minipage}
\end{table}
\setlength{\tabcolsep}{6pt}

\end{document}